\documentclass[useAMS,usedcolumn,usenatbib,usegraphicx]{mn2e}

\usepackage{epsfig,graphicx,natbib}
\usepackage{./mycite}
\usepackage{amssymb}
\usepackage{amsfonts}
\usepackage{amsmath}
\usepackage{color}
 
\newcommand{\ud}{\,\mathrm{d}}

\begin{document}
\title[Weak Lensing and Modified Gravity]{Constraining Modified Gravity and Growth with Weak Lensing
\thanks{Based on observations obtained
 with {\sc MegaPrime/MegaCam}, a joint project of CFHT and CEA/DAPNIA, at
the Canada-France-Hawaii Telescope (CFHT) which is operated by the
National Research Council (NRC) of Canada,
the Institut National des Sciences de l'Univers of the Centre National de
la Recherche Scientifique (CNRS) of France, and the University of Hawaii.
This work is based in part
on data products produced at {\sc Terapix} and the Canadian Astronomy Data
 Centre as part of the Canada-France-Hawaii Telescope Legacy Survey, a collaborative project of
    NRC and CNRS.
}}

\author[Thomas et al.]
{Shaun A. Thomas$^{1}$,
Filipe B. Abdalla$^{1}$,
Jochen Weller$^{1,}$$^{2,}$$^{3}$ \\
$^{1}$Department of Physics and Astronomy, University College London,
Gower Street, London, WC1E 6BT, UK.\\
$^{2}$Universit\"atssternwarte, M\"unchen, Scheinerstr. 1, 81679
M\"unchen, Germany.\\
$^{3}$Cluster of Excellence 'Origin and Structure of the Universe', Boltzmannstr.2, 85748
Garching, Germany.\\
\\
Email: sat@star.ucl.ac.uk
}

\maketitle

\begin{abstract}
The idea that we live in a Universe undergoing a period of acceleration is a new, yet strongly held, notion in cosmology. As this can, potentially, be explained with a modification to General Relativity we look at current cosmological data with the purpose of testing aspects of gravity. Firstly we constrain a phenomenological model (mDGP) motivated by a possible extra dimension. This is characterised by a parameter $\alpha$ which interpolates between $\alpha=0$ (LCDM) and $\alpha=1$ (the Dvali Gabadadze Porrati (DGP) model). In addition, we analyse general signatures of modified gravity given by the growth parameter $\gamma$ and power spectrum parameter $\Sigma$.  We utilise large angular scale ($\theta >$ 30 arcminutes) Weak Lensing data (CFHTLS-wide) in order to work in the more linear regime and then add, in combination, Baryon Acoustic Oscillations (BAOs) and Supernovae. We subsequently show that current weak lensing data is not yet capable of constraining either model in isolation. However we demonstrate that even at present this probe is highly beneficial, for in combination with BAOs and Supernovae we obtain $\alpha < 0.58$ and $\alpha < 0.91$ at $1\sigma$ and $2\sigma$, respectively. Without the lensing data no constraint is possible. Inclusion of all angular scales ($1 \le \theta \le 230$ arcminutes) allows a noticeable but not significant increase in the constraining power of the joint probes with $\alpha < 0.56$ and $\alpha < 0.86$ also at $1\sigma$ and $2\sigma$, respectively. Both analyses correspond to a disfavouring of the flat DGP braneworld model ($\alpha = 1$) at over $2\sigma$. We highlight these are insensitive to potential systematics in the lensing data such as an underestimation of the shear at high redshift. For the growth signature $\gamma$ we show that, in combination, these probes do not yet have sufficient constraining power. Finally, we look beyond these present capabilities and demonstrate that Euclid, a future weak lensing survey, will deeply probe the nature of gravity. A $1\sigma$ error of 0.104 is found for $\alpha$ ($l_{\mathrm{max}} = 500$) whereas for the general modified signatures we forecast $1\sigma$ errors of 0.045 for $\gamma$ and 0.25 for $\Sigma_{0}$ ($l_{\mathrm{max}} = 500$), which is further tightened to 0.038 for $\gamma$ and 0.069 for  $\Sigma_{0}$ ($l_{\mathrm{max}} = 10000$).
\end{abstract}


\section{Introduction}

General relativity, a cornerstone of physics, is arguably one of our greatest intellectual achievements. It is not only elegant and physically motivated, but it makes a whole host of predictions including gravitational waves, the anomalous precession of Mercury and the deflection of light--all of which have been verified. 

Further still, due to its weak yet accumulative nature, gravitation is a force of the large scale. And so cosmology, being the study of the Universe, is a field performed within the formalism of this theory. It is in this way that Cosmology itself becomes yet a further test for gravity. Considering this, today's cosmologists have been posed the most tantalising problem: given that recent precision data from Supernovae, the Cosmic Microwave Background and large scale structure (E.g. \citet{Astier06}, \citet{Dunkley08} and \citet{Percival07}) all indicate that the Universe is undergoing a period of cosmic acceleration, do we treat this as evidence that Einstein's theory of gravitation, or the manner in which we implement it, is incomplete? Or, more conventionally, should we stand by this successful theory and invoke some new matter component--Dark Energy--to explain it? 

Here we study the former and investigate the idea that General Relativity is not general enough. We do not attempt to motivate a new theory of gravity itself but instead aim towards testing existing theories and aspects of general theories with current and future data. In Section~\ref{sec:modifiedgravity} we review the concept of modified gravity, including a generalised braneworld model \citep{DvaliTurner03} we go on to constrain, and touch upon some of its interesting features. One example feature, and thus potential signature of modified gravity, concerns the growth of structure. In Section~\ref{sec:growth} we look deeper at this growth characteristic and attempts to generally parameterise it analogous to the equation of state for dark energy. We highlight how this extra richness in modified gravity can break the observational degeneracy with dark energy models and discuss the ensuing limitations. In addition, we note how modified gravity alters the relationship, relative to GR, between the power spectrum of the potentials and the matter power spectrum which is implicit within weak lensing. Section~\ref{sec:weaklensingasacosmologicalprobe} introduces weak gravitational lensing and its particular importance to modified gravity given that it is sensitive to the expansion history, the growth of structure and the power spectrum. In this section we also detail the survey and data that we use (CFHTLS-wide: \citet{Fu08} - From here on F08) and follow with a discussion of the working caveats, including non-linearities, and how this data in particular circumvents the issue. Section~\ref{sec:lensing} subsequently contains the analysis and constraints of the general braneworld-like model and parameterisation of growth through lensing. It is promptly followed in Section~\ref{sec:BAO} by the addition of BAO and Supernovae data to improve upon these constraints and break parameter degeneracies. For all the analyses in Section~\ref{sec:lensing} and Section~\ref{sec:BAO} we implement a Monte Carlo Markov Chain (MCMC) approach with CosmoMC\footnote{http://cosmologist.info/cosmomc/} \citep{Lewis02} where the resulting plots have been produced with CosmoloGUI\footnote{http://www.sarahbridle.net/cosmologui/}. In Section~\ref{sec:systematics} we briefly highlight potential systematics in the data and quantify any effect on the constraints. We also look beyond present day constraints on gravity in Section~\ref{sec:Euclid} and see how the highly exciting future weak lensing probe Euclid (\citet{Refregier08} and \citet{Cimatti08}) will be able to distinguish between General Relativity and other models of cosmic acceleration and growth. We finish in Section~\ref{sec:conclusion} with a summary of the paper including a discussion of the caveats and limitations as well as suggestions for future work.

\section{Modified Gravity} \label{sec:modifiedgravity}

General relativity itself is a modification of gravity. It superseded the previous established theory of gravity, Newton's Law of Gravitation, with a breathtaking physical principle and mechanism for gravitation phenomena. Although an enormous and elegant change in how we think about gravity it was quite simply necessary: The previous gravitational framework did not explain all gravitational processes. For example, it did not account for the anomalous precession of Mercury. Given the success of Newton's theory many attempts were made to understand this within the existing framework. In fact, even a form of dark matter was invoked in the form of an unseen planet to cause the required procession. 

Today one may argue we face a similar choice with the evidence of accelerated expansion in the Universe. Again early attempts have tried to incorporate some new matter component within the formalism of our current theory of gravity. The simplest procedure has been to introduce a cosmological constant--perhaps arising from vacuum energy--to the usual Einstein field equations. However, a disagreement of 120 orders of magnitude represents a severe fine tuning problem. Other similar avenues have included the introduction of a dynamical scalar field which is either trapped within a false vacuum or slowly rolling down a potential (E.g. \citet{Wetterich87}, \citet{Peebles88}, \citet{Frieman95}, \citet{Ferreira98} and \citet{Albrecht00}). These Quintessence or dark energy models can potentially lead to the desired acceleration.

Alternatively the more radical, but historically successful, route is with another modification to gravity. Starting from the assumption that any viable theory is described by a Langrangian one might at first consider adding terms to the Ricci scalar (R) in the Einstein-Hilbert action for General Relativity given below,

\begin{equation} \label{eq:GRaction}
I_{G} \equiv -\frac{1}{16\pi G} \int \sqrt{-g} R \ud^{4}x.
\end{equation}

In fact this procedure was performed some time ago in the context of inflation (\citealt{Starobinskii80}) but has more recently been analysed for the late-time low curvature universe, in e.g. \citet{CarrollDuvvuriTrodden04}, where the term $1/R$ was added. It was found to have the desired effect of acceleration but is ultimately unfeasible as a realistic alternative due to its failure to comply with solar system constraints (\citealt{ChibaSmithErickcek06}). A generalised modification could assume that the correct underlying theory must be some general function of the Ricci scalar. These models, called f(R) models, are being studied extensively in the literature. One could potentially generalise this even further to functions of the Ricci tensor $R_{\mu \nu}$ and  curvature tensor $R_{\mu \nu \rho \sigma}$, resulting in f(R,$R_{\mu \nu} R^{\mu \nu}$,$R_{\mu \nu\rho \sigma } R^{\mu \nu \rho \sigma}$) gravity. However, this general gravity suffers from higher order instabilities, through Ostrogradski's theorem \citep{Woodard07}, and so analysis has tended to focus on f(R). It is not exclusively this subset of theoretical space that suffers theoretical problems however. Healthly theories of gravity seem to be particularly rare with most suffering from a whole host of theoretical afflictions, from ghost negative energy states to tachyonic behaviour \citep{DurrerMaartens07}. It is also worth noting that if any of these remaining problem-free theories are successful in explaining the observed acceleration we are also faced with the challenge of explaining why vacuum energy, mentioned earlier, does \emph{not} gravitate!

Beyond generalising the function in the Lagrangian we could also look to higher dimensional models. Within the context of cosmology this braneworld scenario can somewhat be described as string theory inspired. For example, normal matter might be confined to a 4-dimensional brane, where the conservation equation $\dot{\rho} + 3H(\rho+p)= 0$ holds firm, but gravity is free to roam into a higher dimensional bulk. One particular braneworld scenario, the Randall-Sundrum model (\citealt{Randall99}), approximates to standard gravity at low energies but becomes five dimensional for high energy and small length scales. For late time acceleration however we desire something that will change over large distances and low energy scales. The DGP model\footnote{See \citet{Lue06} for an extensive review. } of Dvali-Gabadadze-Porrati  (\citealt{DvaliGabadadzePorrati00}), described by the Lagrangian in Equation \eqref{eq:DGPaction}, is exactly this. 

\begin{equation} \label{eq:DGPaction}
I_{G} \equiv \frac{-1}{16 \pi G} \Big[ \frac{1}{r_{c}} \int_{\mathrm{bulk}} \! \! \ud^{5}x \sqrt{-g^{(5)}}R^{(5)} + \int_{\mathrm{brane}} \! \! \ud^{4} x \sqrt{-g}R\Big]
\end{equation}

It was originally created consisting of a 4-dimensional Minkowski brane within a 5-dimensional Minkowski bulk and with no motivation towards explaining dark energy. However the generalisation (\citealt{Deffayet00}) to a Friedmann-Robertson-Walker brane gave rise to a self-accelerating solution. Gravity leaking from this 4D brane into the bulk over large scales gives rise to the acceleration through a weakening effect. The resulting Friedmann equation is a modification to the General Relativistic case and is given by Equation \eqref{eq:DGPFriedman} with $r_{c}$, the cross over scale, specified in Equation \eqref{eq:rc}.

\begin{equation} \label{eq:DGPFriedman}
H^{2} - \frac{H}{r_{c}} = \frac{8 \pi G \rho}{3}
\end{equation}

\begin{equation} \label{eq:rc}
r_{c} = \frac{1}{H_{0} (1- \Omega_{m})}
\end{equation}

With this modification we have a full description of the expansion history. This also allows us to work towards understanding the growth of large scale structure giving two observational signatures that would enable us to carry out a cosmological study. The difference in the background acceleration itself is enough to produce a difference in the growth of structure. This can be seen in the second term, the Hubble drag, in the growth of density perturbations $\delta$, for GR below,

\begin{equation} \label{eq:standardgrowth}
\ddot{\delta} + 2H\dot{\delta} = 4\pi G \rho_{m} \delta.
\end{equation}

However, assuming that the only alteration is via changes in $H$ is fortunately incorrect. It is fortunate because it is the extra modification that allows us to break the degeneracy between some general dark energy within GR, which can replicate any desired expansion history, and this modified gravity model (this will be addressed in more detail in Section~\ref{sec:growth}). The correct approach regarding the evolution of perturbations in this gravitational framework is particularly difficult and was tackled by \citet{LueScoccimarroStarkman04} and then by \citet{KoyamaMaartens06}. It was found that treating gravity as 4-dimensional, which leads to Equation \eqref{eq:standardgrowth}, induces an inconsistency in the 4-dimensional Bianchi identities. Instead with the full and complete five-dimensional analysis, and assumptions of a quasi-static regime and sub-horizon scales, they found the metric perturbations on the brane to be,

\begin{equation} \label{eq:phiDGP}
k^{2} \phi = -4 \pi G a^{2} (1 - \frac{1}{3 \beta}) \rho_{m} \delta
\end{equation}

\begin{equation} \label{eq:psiDGP}
k^{2} \psi = -4 \pi G a^{2} (1 + \frac{1}{3 \beta}) \rho_{m} \delta
\end{equation}

\begin{equation} \label{eq:beta}
\mathrm{where} \qquad \beta = 1 - 2r_{c} H (1 + \frac{\dot{H}}{3H^{2}}).
\end{equation}

It is this $\beta$ factor within Equation \eqref{eq:psiDGP} that breaks the expansion degeneracy and modifies Equation \eqref{eq:standardgrowth} becoming rather,

\begin{equation} \label{eq:DGPgrowthdifferential}
\ddot{\delta} + 2H\dot{\delta} = 4\pi G \Big(1+ \frac{1}{3 \beta}\Big) \rho_{m} \delta.
\end{equation}

\begin{figure}
      \includegraphics[width=3.5in,height=3.5in]{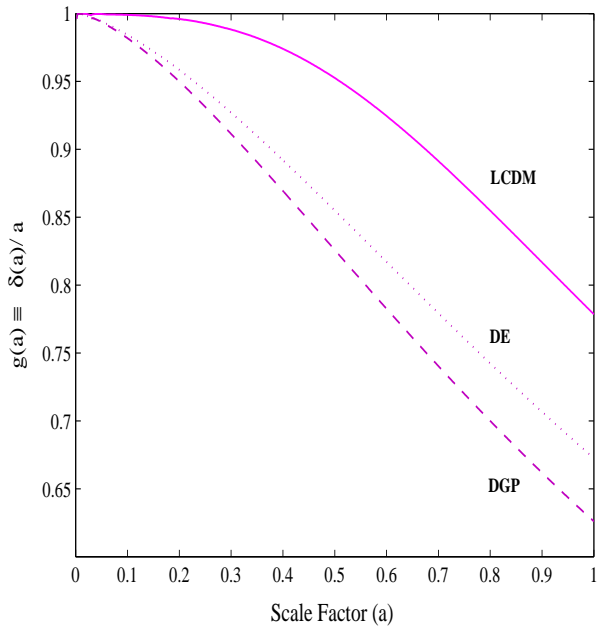}
    \caption{\small{$\mathrm{g(a)} \equiv \delta(a)/a$, the linear growth, is plotted for a range of late time acceleration models. The solid line demonstrates the growth for LCDM, the dashed for the five-dimensional braneworld model DGP and the dotted for a dark energy model with identical expansion history to DGP ($w_{0} = -0.78$ and $w_{a} = 0.32$ where $w(a) = w_{0} + (1-a) w_{a}$). The difference in the expansion history seems to give a significant suppression in growth relative to a pure cosmological constant. The effect, however, of the five-dimensional perturbation analysis not only adds to the suppression for DGP but breaks the degeneracy between itself and the attempt at replication from the smooth dark energy model.}}
    \label{fig:growths}
\end{figure}

One can see the effect of this analysis by looking at Figure \ref{fig:growths}. We have plotted the linear growth factor for LCDM, DGP and a dark energy model with the same expansion history as DGP. It is evident that the expansion history has considerable influence on the linear growth of structure with a suppression in the dark energy model relative to a cosmological constant. The effect of the five-dimensional perturbation analysis adds to this suppression and acts to clarify the deviation between the dark energy and DGP model.

With the modified Friedman equation, the correct linear growth equation and several caveats it is now possible for one to perform tests on the expansion history and/or large scale structure for this particular modification to gravity. Some of these tests already exist and it has been found that DGP is under tension from the recent influx of cosmological data (E.g. \citet{SongSawickiHu07} and references therein). It is also worth noting that this model is potentially not without some of the theoretical problems alluded to above with notions of a ghost (\citet{Koyama05} and \citet{Gorbunov06}) and a strong coupling problem (\citealt{Rubakov03}). 

We do, however, go beyond DGP as an isolated theory and examine a phenomenological model that is motivated by the concept of an extra dimension with infinite extent. This model, first introduced by \citet{DvaliTurner03}, interpolates between LCDM and DGP and allows for the presence of the extra dimension through corrections to the Friedmann equation with the addition of the parameter $\alpha$ shown in Equation \eqref{eq:mDGPFriedman} and $r_{c}$ in Equation \eqref{eq:rcmDGP}.

\begin{equation} \label{eq:mDGPFriedman}
H^{2} - \frac{H^{\alpha}}{r_{c}^{2-\alpha}} = \frac{8 \pi G \rho}{3}
\end{equation}

\begin{equation} \label{eq:rcmDGP}
\mathrm{where} \qquad r_{c} = (1-\Omega_{m})^{\frac{1}{\alpha-2}}H_{0}^{-1}.
\end{equation}
\noindent
It is clear that in this case LCDM is recovered when $\alpha = 0$ and DGP when $\alpha=1$. Furthermore, it is worth noting that $\alpha < 0$ leads to effective equation of states less than $-1$, whereas $\alpha \gtrsim 1$ acts to disrupt both the long matter era needed for structure formation and the limits set by Big Bang Nucleosynthesis (BBN), while $\alpha > 2$ corresponds to early universe braneworld modifications. Using Equation \eqref{eq:mDGPFriedman} it is possible to test the model with probes of the expansion history and indeed this has already been performed by \citet{YamamotoBassettNichol06} with Baryon Acoustic Oscillations. If one wants to go beyond this and include tests of large scale structure then one needs a formalism for the growth of density perturbations analogous to Equations \eqref{eq:psiDGP}, \eqref{eq:beta} and therefore \eqref{eq:DGPgrowthdifferential}. The problem in this scenario is that in order to deduce the growth of perturbations one needs an underlying covariant theory and all that exists in this modified DGP model (mDGP) is a parameterisation. \citet{Koyama06} circumvented this obstacle by taking DGP (a limit in the model) as the structure of the theory. It was subsequently found that the metric perturbations take the same form as Equations \eqref{eq:phiDGP} and \eqref{eq:psiDGP} but instead with,

\begin{equation} \label{eq:mDGPbeta}
\beta = 1-\frac{2}{\alpha} (H r_{c})^{2 - \alpha}\Big(1 + \frac{(2-\alpha)\dot{H}}{3 H^{2}}\Big).
\end{equation}

\begin{figure}
      \includegraphics[width=3.5in,height=3.5in]{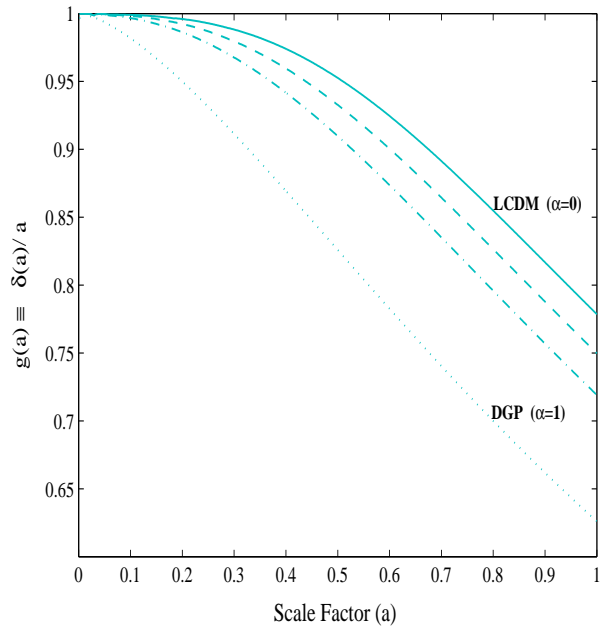}
    \caption{\small{$\mathrm{g(a)} \equiv \delta(a)/a$, the linear growth, is plotted for various values of $\alpha$ that characterise the phenomenological general braneworld-like (mDGP) model. The solid line demonstrates the growth for $\alpha=0$ (LCDM), the dashed for $\alpha=0.25$, the dash-dotted for $\alpha=0.5$ and the dotted for $\alpha=1$ (DGP). Once again it is evident that the more DGP-like end of the $\alpha$ spectrum experiences more suppression in the growth of density perturbations.}}
    \label{fig:alpha_growths}
\end{figure}

Figure \ref{fig:alpha_growths} demonstrates how the growth of density pertubations alter within the mDGP model - from LCDM ($\alpha=0$) to DGP ($\alpha=1$). As in the previous Figure it is clear that there is a suppression of growth at the more DGP end of the $\alpha$ spectrum.

Although a parameterisation of a general large extra dimensional model the mDGP model now has a definite Friedmann equation that governs the expansion history and, assuming the structure of any underlying theory, it has a set of metric perturbation equations and a corresponding density perturbation equation. We can therefore treat this as a specific model which we choose to constrain later in the paper. It is worth noting that using this as a measure of deviation from General Relativity or as a parameterisation of general modified gravity is not our aim. This would constitute a poor choice of parameter given the severe lack of generalness and incompleteness with regards to the concept of modified gravity as a whole. We do however touch upon the idea of parameterising modified gravity in the next section. This model has, however, been extremely illustrative with regards to the extra richness that can occur in modified gravity. Not only does it have varying expansion histories but potentially a whole range of perturbation equations which alters the growth of structure and the relationship to the power spectrum. This is particularly useful when attempting to distinguish between LCDM, general dark energy and modified gravity, and insightful to the probes that will be most adept at detecting them. Again, we discuss these issues in the following sections.

\section{Growth} \label{sec:growth}

The alteration in the growth of structure within the mDGP model demonstrated an additional observational characteristic that allows us to further constrain the model and potentially break the degeneracy with an expansion replicating dark energy. It also highlights the possibility of searching for signatures of modified gravity in current data by looking for changes in the growth of structure. However, even if we do have a plethora of new and viable physical principles that explain dark energy naturally we might want a way to quantify how these models affect this growth signature. If we do not have the models we might just want to test the influx of data for \emph{hints} of something unusual. Either way it is possible to parameterise this extra growth. 

This notion of parameterising growth is analogous to the familiar parameterisation of the background expansion into $w_{0}$ and $w_{a}$. This is sufficient in describing and restricting the multitude of possible dark energy models and expansion histories. It is now common procedure to examine data and convert it into constraints on various cosmological parameters including $w_{0}$ and $w_{a}$. We might therefore like to extend this parameter space and allow for the additional signatures of gravity. One possible parameterisation for growth is given by $ \gamma $ in Equation \eqref{eq:lindergamma} and was first introduced by \citet{Peebles08} and  \citet{Lahav91} and later discussed in \citet{Wang98}, \citet{linder05}, \citet{HutererLinder07} and \citet{linderCahn07}.

\begin{equation} \label{eq:lindergamma}
\frac{\delta}{a} \equiv g(a) = \mathrm{exp}\Big(\int_{0}^{a}  (\Omega_{\mathrm{m}}(a)^{\gamma} - 1 ) \ud \mathrm{ln} a\Big)
\end{equation}

By once again looking at Figure \ref{fig:growths} we can see that the growth factor g(a) is affected by the expansion history and by the gravitational framework. It is worth noting that the $\gamma$ parameterisation distinguishes the two contributions to the growth, encapsulating the latter in isolation. This is due to the impact from the expansion being absorbed into $\Omega_{m}(a)$ thus leaving $\gamma$ to pick out any remaining remaining contribution. It is in this way that $\gamma$ has become known as a modified gravity or beyond-Einstein parameter. It is easy to see why given that it detects changes to the growth not associated with expansion. This could be down to a change in the force law acting on matter represented, for example, by the extra factor in Equation \eqref{eq:DGPgrowthdifferential}. And as we alluded to earlier, evident in Figure \ref{fig:growths}, this allows us to distinguish between dark energy and modified gravity. However, as highlighted in \citet{KunzSapone07} there exists an interesting caveat. They found that contrary to Figure \ref{fig:growths} one could force some arbitrary and generic dark energy to replicate the growth of DGP. This was achieved by allowing for dark energy models with lower sound speeds ($c_{s}^{2} \neq 1$) which in turn induce clustering in the fluid. The clustering instigates a deepening of the gravitational potential wells thus leading to a magnification in the metric perturbations and subsequently an increase in the growth. In addition the existence of anisotropic stress was permitted which had the effect of suppressing growth. With a careful balance between stress and sound speed they succeeded in replicating g(a) for DGP. Now although highly fine tuned it is worth keeping in mind that observationally detecting some non-LCDM growth factor, or $\gamma$, would therefore not necessarily constitute modified gravity. In this way, unless one allows for only smooth non-clustering dark energy the growth parameter is not just a modified gravity parameter. Instead it has the ability to pick up on clustered dark energy and modified gravity both of which would be interesting. Given the above it is therefore our intention to test the growth parameter and see whether current data or a future probe can pick out this subtle but potentially important effect. Indeed there exists a few early attempts, including constraints from peculiar velocity measurements from low redshift Supernovae (\citealt{Abate08}) as well as future survey forecasts from  \citet{AmendolaKunzSapone07}, \citet{HutererLinder07} and \citet{HeavensKitchingVerde07}.

\begin{figure}
      \includegraphics[width=3.5in,height=3.5in]{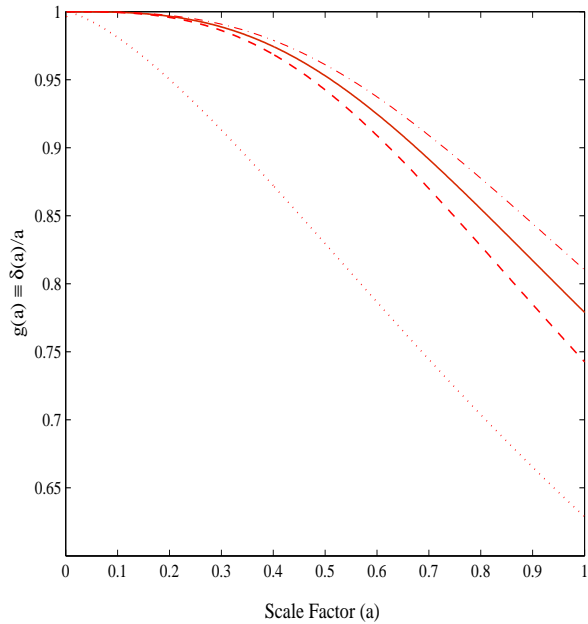}
    \caption{\small{$\mathrm{g(a)} \equiv \delta(a)/a$, the linear growth, is plotted for various values of $\gamma$ the parameterisation of growth resulting potentially from a change in force law. The solid line represents the growth for the LCDM paradigm ($w_{0} = -1, \quad w_{a}=0$) with the corresponding growth parameter of $\gamma = 0.55$. The dashed line shows the growth for $\gamma=0.68$ which is the same as resulting from flat DGP but with the same expansion as in LCDM. The dotted line also has $\gamma=0.68$ but now with $w_{0} = -0.78$ and $w_{a} = 0.32$ thus completely specifying the growth of the example DGP model. Finally the dot-dashed line shows the growth for a LCDM expansion but $\gamma=0.45$. It is quite clear that a high value of the growth parameter corresponds to a suppression of growth. This potentially arises from a weakening of gravity.
 }}
    \label{fig:gamma_growths}
\end{figure}

Figure \ref{fig:gamma_growths} demonstrates the result of varying this parameter on the linear growth factor. The growth for standard LCDM corresponds to $\gamma = 0.55$ whereas for flat DGP it corresponds to $\gamma=0.68$. In this way it is clear that a higher growth parameter results in a suppression of growth. 

It is worthwhile noting that other attempts at parameterising modified gravity have been made which aspire to encapsulate the properties of gravity similar to the Parameterised Post-Newtonian (PPN) parameters for local gravity constraints (\citealt{Will93}). For example, these include parameterising the relationship between the two metric potentials ($\phi$ and $\psi$) and/or quantifying any modification to the Poisson equation (E.g. \citet{AmendolaKunzSapone07}, \citet{HuSawicki07b}, \citet{IshakUpadhyeSpergel05}, \citet{Jain07}, \citet{Daniel08} and \citet{BertschingerZukin08}). In fact these observations lead us on to the final modified gravity signature and characteristic we must consider before our analysis with weak lensing. For any deviation in the Poisson equation or between the metric potentials causes a deviation in the relationship between the power spectrum of the potentials and the power spectrum of the density contrast. Failure to account for this by not modifying the corresponding lensing equations will render any analysis incomplete. To understand this we use the notation of \citet{AmendolaKunzSapone07} but in doing so note the equivalence of their $Q$ and $\eta$ (defined below) to $\mathrm{\tilde{G}_{eff}}$ and $\eta$ given in the other thorough consideration by \citet{Jain07}.

Firstly, as will be further detailed in the following section, let us notice that in weak lensing, our probe of choice, the deflection of light is sensitive to the sum of the metric potentials $\phi + \psi$. Therefore, we require the power spectrum within the lensing statistic (Equation \eqref{eq:convergencepowerspectrum}) to be the power spectrum of $\phi + \psi$, written $P_{(\phi + \psi)}$, which we then relate to the matter power spectrum $P_{\delta}$. Defining the matter power spectrum in Equation \eqref{eq:deltapowerspectrum} and the potential power spectrum similarly it is obvious that a general relationship between $P_{\delta}$ and $P_{(\phi + \psi)}$ relies on the relationship between $\phi + \psi$ and $\delta$ which in turn depends on the Poisson equation and the relationship between $\phi$ and $\psi$. This is where the parameters $Q$ and $\eta$ (or equivalently $\mathrm{\tilde{G}_{eff}}$ and $\eta$) are particularly useful. Here $Q$ parameterises any modification in the Poisson equation relating the metric variable $\phi$ to the matter density $\rho$. $\eta$ on the other hand, also defined below, describes the relationship between $\phi$ and $\psi$. 

\begin{equation} \label{eq:poissonequation}
k^{2} \phi = - 4 \pi G Q \rho a^{2} \delta
\end{equation}
\begin{equation} \label{eq:etadefintion}
\psi \equiv (1 + \eta) \phi
\end{equation}
\noindent
Now adding the two metric potentials together, substituting $\psi$ for $\eta$ and $\phi$ and transforming the combined Poisson equation, as given in the definition of the matter power spectrum below

\begin{equation} \label{eq:deltapowerspectrum}
<\delta(\vec{k_{1}},z) \delta(\vec{k_{2}},z)> = (2 \pi)^{3} \delta(\vec{k_{1}}+\vec{k_{2}}) P_{\delta}(k,z),
\end{equation}
\noindent
and similarly for the potential, it is possible to see the general relationship between the power spectra,

\begin{equation} \label{eq:powertopower}
P_{(\phi + \psi)}(k,z) = \frac{ (8 \pi G)^{2} \rho^{2} a^{4} [Q(1+\frac{\eta}{2})]^{2} P_{\delta}(k,z)}{k^{4}}.
\end{equation}

We can, following the notation of \citet{AmendolaKunzSapone07}, define $\Sigma \equiv Q(1 + \frac{\eta}{2})$, giving the modification to the power spectrum more succinctly as,

\begin{equation} \label{eq:powertopowersigma}
P_{(\phi + \psi)}(k,z) = \frac{ (8 \pi G)^{2} \rho^{2} a^{4} \Sigma^{2} P_{\delta}(k,z)}{k^{4}}.
\end{equation}

In a standard cosmological scenario, such as LCDM for example, one will have $\eta = 0$ and $Q = 1$ rendering $\Sigma = 1$. This results in the standard relation between the power spectra as often assumed in the literature. It is clear therefore that neglecting $\Sigma$ is tantamount to constraining the subset of modified gravity models that do not alter the power spectrum relation from the General Relativistic case. Surprisingly the mDGP model studied earlier is one such model. This can be seen by simply adding Equations \eqref{eq:phiDGP} and \eqref{eq:psiDGP} together and observing the cancellation in $\beta$. Therefore, for this model $\Sigma = 1$ also and as a result just the expansion history and growth will be altered. Generally however if one strives to include general models $\Sigma$ should be allowed to vary. Note that it acts to modify the amplitude of the power spectrum and so a constant value is degenerate with $\sigma_{8}$. Accordingly, as introduced in \citet{AmendolaKunzSapone07}, we consider the general parameterisation given by,

\begin{equation} \label{eq:sigma}
\Sigma(a) = 1 + \Sigma_{0} a.
\end{equation}

It is our intention therefore to constrain and forecast for the specific mDGP gravity model and then, separately, constrain the general characteristics of modified gravity. For the latter we choose to set $\Sigma = 1$ (or $\Sigma_{0} = 0$) due to limitations in data and constrain $w_{0}$ and $\gamma$ signatures only. We later include $\Sigma_{0} \neq 0$ for the Euclid forecasts. It is worthwhile noting that $\gamma$ and $\Sigma$ are functions of the more fundamental $Q$ and $\eta$ and that these parameters themselves are, most generally, $Q(k,a)$ and $\eta(k,a)$ \citep{AmendolaKunzSapone07}.

\section{Weak Lensing as a Cosmological Probe} \label{sec:weaklensingasacosmologicalprobe}

The deflection of light by mass is given by the transverse gradient of the metric potentials integrated along the path length,

\begin{equation} \label{eq:angledeflection}
\varphi = - \int \partial(\psi + \phi) \mathrm{d}s.
\end{equation}
\noindent
This acts to not only change the apparent position of some point source but, in turn, distort the shape of distant source galaxies. Indeed, one can relate the observed position of the image $\vec{\theta}_{I}$ to the true position of the source $\vec{\theta}_{S}$, in the plane of the sky, by

\begin{equation} \label{eq:lensequation}
\vec{\theta}_{I} = \vec{\theta}_{S} + \frac{D(\chi_{s} -\chi)}{D(\chi)} \vec{\varphi},
\end{equation}
\noindent
where $D(\chi)$ is given by the comoving angular diameter distance. The subsequent image distortion is given by the differential of this lens equation resulting in the Jacobian

\begin{equation*} \label{eq:jacobian}
\vec{A} = \frac{\mathrm{d} \vec{\theta}_{S}}{\mathrm{d}\vec{\theta}_{I}} = \left(
\begin{array}{cc}
1 - \kappa - \gamma_{1} & \gamma_{2}  \\
\gamma_{2} & 1 - \kappa + \gamma_{1}
\end{array} \right),
\end{equation*}
\noindent
where the convergence $\kappa$ is reconstructed from the shear $\gamma$ (where $\gamma = \gamma_{1} + i\gamma_{2}$) which in turn is measured from galactic ellipticities \citep{Refregier03}. From the point of view of a cosmological model it is the convergence that we are interested in. This convergence, which represents a weighted projected mass distribution on the plane of the sky, is given \citep{Bartelmann01} for a general mass distribution by

\begin{equation} \label{eq:convergence}
\kappa = \frac{3 \Omega_{m} H_{0}^{2}}{2 c^{2}} \int_{0}^{\chi_{s}} \mathrm{d} \chi \frac{D(\chi) D(\chi_{s} - \chi)}{\chi_{s}} (1+z) \delta(\chi).
\end{equation}
\noindent
Given that we wish to analyse this distribution in a statistical way we use the defintion of the power spectrum - analogous to Equation \eqref{eq:deltapowerspectrum} - to find the expression for the convergence power spectrum

\begin{equation} \label{eq:convergencepowerspectrum}
P_{\kappa}(l) = \frac{9 \Omega_{m}^{2} H_{0}^{4}}{4 c^{4}} \int_{0}^{\chi_{H}} \mathrm{d} \chi \Big[ \frac{g(\chi)}{a(\chi)} \Big]^{2} P_{\delta} \Big( \frac{l}{\chi}, \chi \Big),
\end{equation}
\noindent
where the geometric comoving angular diameter distance terms have been absorbed into $g(\chi)$. It is therefore now clear, given our earlier discussion, how weak lensing as a cosmological probe is particularly useful in studies of modified gravity. For this statistic is sensitive to the growth of structure via the presence of the linear growth factor squared ($g^{2}(a)$) in the matter power spectrum $P_{\delta}$. It is sensitive to the expansion history through the terms in the square brackets and through the Hubble drag in the growth of structure and also, as discussed at the end of the last section, it is sensitive to the relation between the power spectrum of the potentials and density. In the equation above the relation from $P_{\phi + \psi}$ to $P_{\delta}$ has already been performed assuming GR as given routinely in the literature. Again, it is worth reiterating that if there is a modification to the Poisson equation and/or to the anisotropic stress one must augment this powerspectrum with the approapriate prefactor given, for example, as in Equation \eqref{eq:powertopower}. In addition to these sensitivities it is worth adding that because the deflection is given by the potentials, which are sourced by mass irrespective of being baryonic or dark, weak lensing does not suffer from any unknown bias. That is, it probes the entirety of the mass distribution.

While this probe, in principle, is excellent for our chosen study it is worthwhile noting that the shear signal is a small $1\%$ distortion on the already existing intrinsic ellipticity. This provides a thorough technical challenge that is being combated with a combination of large galaxy number analyses and refined shear measurement techniques (\citet{Heymans05}, \citet{Massey06} and \citet{Bridle08}). Further still, the first detections of weak lensing are particularly recent (\citet{Bacon00}, \citet{Kaiser00}, \citet{Wittman00} and \citet{vanWaerbeke00}) and so in this way lensing is very much a highly promising, yet developing, cosmological probe. Despite this there are already a number of papers in the literature that have addressed the relationship between weak lensing and modified gravity or dark energy, such as \citet{UzanBernardeau01}, \citet{Schimd05}, \citet{DoreMartigMellier07}, \citet{Schimd07}, \citet{AmendolaKunzSapone07}, \citet{Jain07} and  \citet{Tsujikawa08}. With these studies and potential modified gravity attributes in lensing it is imperitive to realise that there does exist a severe caveat. This is due to the fact that lensing probes into the non-linear regime. We are fortunate to be able to use a fitting function (E.g. \citealt{Smith03}) for the non-linearities in standard gravity. Unfortunately this is poorly understood in any deviation from the current framework and subsequent implementation of the fit would, technically, be invalid. Therefore until N-body simulations have been undertaken that could generalise the fitting function or accurately quantify deviations from it we must strive to work in the linear regime as much as possible. Although some attempts have been made at quantifying the validity of the present fits (E.g. \citet{LaszloBean07} and \citet{Oyaizu08}) we keep, for now, a strict and linear only analysis. There are, in addition, other benefits in avoiding the inclusion of small scales such as the presence of intrinsic ellipticity correlations, shear-shape correlations and the presence of non-Gaussianity in the error. We therefore utilise the data provided by F08 based on the CFHTLS-wide survey which, due to its range of large angular scales (up to 230 arcminutes) probing the more linear regime, is ideal for work on non-LCDM cosmology such as this.

\subsection{CFHTLS} \label{sec:CFHTLS}

The Canada-France-Hawaii Telescope Legacy Survey \footnote{http:/www.cfht.hawaii.edu/Science/CFHTLS/} (CFHTLS), based on the MEGAPRIME/MEGACAM instrument, is an ongoing survey with a target of 450 nights extending over 5 years. The recent analysis by \citet{Benjamin07} has gone beyond the initial releases and investigations by \citet{Semboloni06} and \citet{Hoekstra06} which themselves were successful in deriving constraints on the $\Omega_{m}- \sigma_{8}$ degeneracy and demonstrating the evolution of the shear signal with redshift. This was achieved in \citet{Benjamin07} through a better understanding of the redshift distribution and having an increased area. This, while marking significant progress, is still not the most optimal lensing analysis for this work. This is because they are potentially sensitive to the growth of structures on non-linear scales which, as we emphasised above, is undesirable for a current study of beyond-Einstein cosmology and weak lensing. We therefore look to the 3rd year CFHTLS-wide release (T0003) given by \citet{Fu08} (F08). Although having a smaller field of view than \citet{Benjamin07} it utilises much larger angular scales (into the linear regime) also avoiding many of the potential systematics mentioned at the end of the last section. It is because of this that both works reveal approximately equivalent cosmological constraints and little constraining power is lost. 

\begin{figure}
      \includegraphics[width=3.5in,height=3.5in]{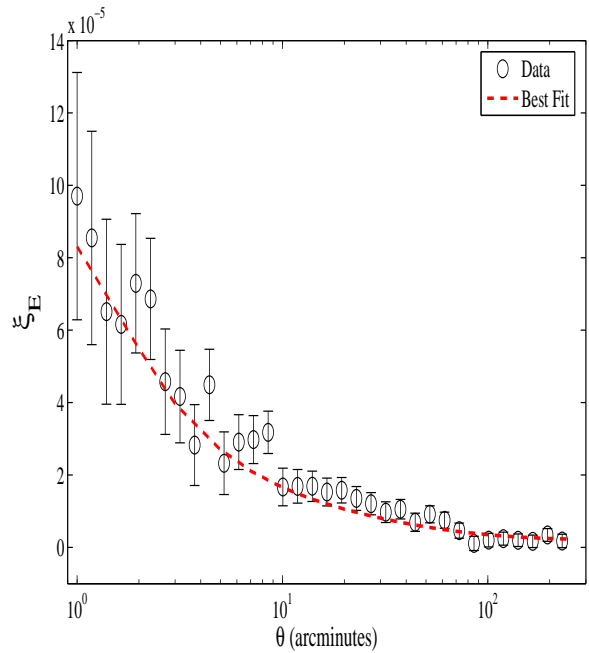}
    \caption{\small{The open circles with associated errorbars represent the $\xi_{E}$ two-point statistic as a function of $\theta$ (arcminutes) for the CFHTLS-wide survey used in this paper. We however selectively use scales greater than 30 arcminutes to remove the unknown non-linear effects and in doing so strive towards a more linear analysis. The red dashed line shows the best fit values as found with the combined probes mDGP analysis (Section~\ref{sec:BAO}).}}
    \label{fig:paper_E_correlation}
\end{figure}

The current sky coverage of $57 \mathrm{deg^{2}}$, approximately $35\%$ of the final CFHTLS target area, is reduced to  $34.2 \mathrm{deg^{2}}$ after masking and the removal of various contaminants. Eventually including five bands this $i'$ band study stretches to a magnitude of $i'_{AB} = 24.5$ and encapsulating nearly 1.7 million galaxies has an effective galaxy number density of $n = 13.3 \mathrm{gal/arcmin^{2}}$. The data (F08) comes in the form of several two point statistics which are relevant to this study. We choose to utilise the E correlation function which is shown in Equation \eqref{eq:Ecorrelation} and displayed along with the cosmological best fit in Figure \ref{fig:paper_E_correlation}.

\begin{equation} \label{eq:Ecorrelation}
\xi_{E} = \frac{1}{2\pi} \int_{0}^{\infty}  l P_{\kappa}(l) J_{0}(l \theta) \ud l
\end{equation}

As for the aperture mass $< \! M^{2}_{\mathrm{ap}} \!>$ and shear top hat variance $<|\gamma|^{2}>$ two point statistics this is a weighted transform of the convergence power spectrum. In this case it is given by a zeroth order Bessel function of the first kind $J_{0}$. It is in this way that the two point functions vary in their sensitivity to various aspects of the power spectrum and in turn any systematics. $\xi_{E}$ suffers from a constant offset resulting from a mixing of E and B-modes. A finite survey size introduces a maximum angular scale which prevents a complete calculation of the shear correlation function over larger ranges. This is needed for a separation of E and B \citep{Kilbinger06}. To alleviate this we alter the statistic $\xi_{E}$ to $\xi_{E} + c'$ including the constant offset $c'$ as an extra parameter. An expression is then obtained for the offset which represents the best fit offset ($\mathrm{d} \chi^{2}/\mathrm{d} c' = 0$) for each parameter choice. This constitutes an analytic marginalisation over $c'$ \citep{Lewis02}. We subsequently find the expression to be

\begin{equation} \label{eq:offsetmarginalisation}
c' = \frac{\displaystyle{\sum_{i,j} (C^{-1})_{ij}
(\xi_i - D_i)}}
{\displaystyle{
\sum_{i,j}(C^{-1})_{ij}}},
\end{equation}
\noindent
 where the element $\xi_{i}$ is the model correlation function, $D_{i}$ the data and $C$ the full covariance matrix between all the elements. Furthermore, the B correlation function $\xi_{B}$ which describes the curl component of the shear field, as opposed to $\xi_{E}$ which measures the curl-free component, is expected to be non-zero only for non-lensing contributions to the shear \citep{Crittenden02}. It is because of this that $\xi_{B}$ is an excellent check on any contamination of the lensing signal. F08 found no real B-mode contribution except for the presence of a very small signal at large angular scales. They find however that their cosmological conclusions are not affected by this potential mode. In addition, it was shown in F08 that there was no significant deviation in cosmological constraints across any of the aforementioned two point statistics. \citet{DoreMartigMellier07}, who also looked at modified gravity models in the context of this CFHTLS-wide data, came to a similar conclusion. It is worth noting that other cosmological studies of this data set include a phenomenological modified gravity analysis \citep{Daniel08} and more recently an early study of the neutrino mass \citep{Tereno08}. We therefore choose for simply to use only one statistic ($\xi_{E}$).

\begin{figure}
      \includegraphics[width=3.5in,height=3.5in]{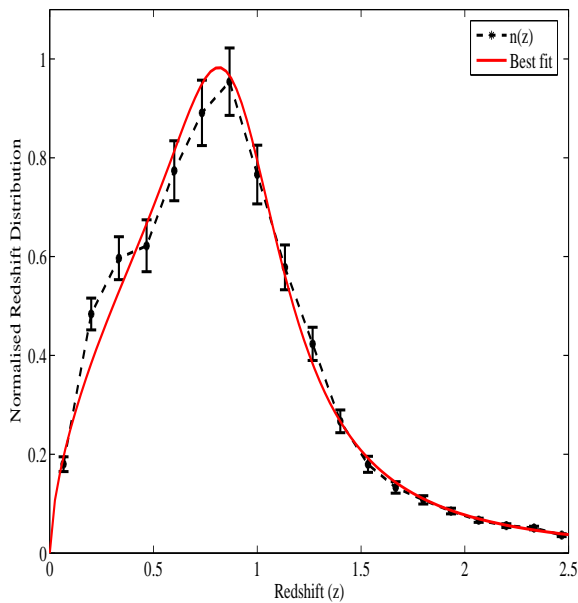}
    \caption{\small{The black dashed line represents the source redshift distribution with associated errors in the bins. The red solid line is given by the fitting function in Equation \eqref{eq:FU_n_z}. The fit is drawn for the function evaluated at the best-fit points as deduced by the combined probes analysis for mDGP (Section~\ref{sec:BAO}). This corresponded to best fit values: $a = 0.614 \pm 0.034$, $b = 8.11 \pm 0.681$, $c = 0.627 \pm 0.0610$ and $A = 0.6462$ consistent with F08.}}
    \label{fig:galaxydistribution}
\end{figure}

The redshift distribution of the source galaxies, which weak lensing is critically sensitive to, has been calibrated for the CFHTLS study using \citet{Ilbert06}. We decide to follow F08 and model this redshift distribution using the function
\begin{equation} \label{eq:FU_n_z}
n(z) = A \frac{z^{a} + z^{ab}}{z^{b} + c} \quad \mathrm{where} \quad  A = \Big(\int_{0}^{z_{\mathrm{max}}}\frac{z^{a} + z^{ab}}{z^{b} + c} \ud z \Big)^{-1},
\end{equation}
\noindent
where $A$ is the normalisation and a, b and c are three extra parameters to be varied and marginalised over in the cosmological fit. They found that Equation \eqref{eq:FU_n_z} enables a closer fit to the distribution data that other common n(z) fitting formulae. The observed normalised redshift distribution, along with the fitting function evaluated at the best fit points found in the mDGP combined probes cosmological run (Section~\ref{sec:BAO}), is shown in Figure \ref{fig:galaxydistribution}.

\section{Constraints}

\subsection{Lensing} \label{sec:lensing}

Having earlier discussed the characteristics of certain late-time acceleration models in Section~\ref{sec:modifiedgravity} and Section~\ref{sec:growth}, and having chosen a probe that is potentially capable of picking out these particular behaviours with weak lensing, we are now in a position to make cosmological constraints based on the data which we decided, in Section~\ref{sec:CFHTLS}, was the most suitable. That is, we start by making an analysis of the mDGP gravitational model and then separately the parameterisation of growth with the F08 CFHTLS-wide lensing data.

We perform a full likelihood analysis using a Monte Carlo Markov Chain (MCMC) approach on a set of 7 cosmological parameters for the mDGP analysis and 8 for the growth parameterisation analysis. We vary $\Omega_{m}$, $h$, $\sigma_{8}$, $a$, $b$, $c$ which are common to both models, in addition to $\alpha$ for mDGP and $w_{0}$ and $\gamma$ for the growth. The $w_{0}$ is not included for the former model as any given $\alpha$ uniquely specifies its own expansion history as it does for the evolution of perturbations--from GR to DGP. We decide not to vary the spectral index $n_{s}$ because the data needed at large angular separations will be insufficient for any constraint. Instead we set it to $n_{s} = 0.963$ consistent with the best fit five-year WMAP result (\citealt{Dunkley08}). Likewise we neglect varying $w_{a}$ and $\Sigma_{0}$ for the growth model due to limitations in the data. Instead we set them to $w_{a}=0$ and $\Sigma_{0}=0$. Finally, we assume a flat universe throughout.

For the lensing analysis the Gaussian log-likelihood is given by

\begin{equation} \label{eq:lensingchisquared}
\chi^{2} = \frac{1}{2} \sum_{\mathrm{ij}} (D_{\mathrm{i}} - T_{\mathrm{i}})(C^{-1})_{\mathrm{ij}}(D_{\mathrm{j}} - T_{\mathrm{j}}),
\end{equation}
\noindent
where the data vector $\vec{D}$ is given by the measured $\xi_{E} (\theta_{i})$. The theoretical prediction deduced at the corresponding angular scale $\theta$ is represented by entries in $\vec{T}$ and, finally, $C^{-1}$ is the inverse covariance matrix for the chosen observable. At present variation in either $a$, $b$ or $c$ in the redshift distribution is detected implicitely through the alteration of the model matter power spectrum. However, in implementing the MCMC approach regions of parameter space could potentially be sampled that would correspond to configurations of $a$, $b$ and $c$ incompatible with just the knowledge of the redshift distribution alone. We therefore follow the procedure in F08 and multiply the likelihood above by the likelihood of the redshift distribution given by

\begin{equation} \label{eq:nzchisquared}
\chi^{2}_{\mathrm{n(z)}} = \frac{1}{2} \sum_{\mathrm{i}} \frac{(n_{i} - n(z_{i}))^{2}}{\sigma^{2}_{\mathrm{i}}},
\end{equation}
\noindent
where $n_{i}$, the observed number of galaxies in a bin, are shown in Figure \ref{fig:galaxydistribution}. $n(z_{i})$ represent the values of the fitting function in Equation \eqref{eq:FU_n_z} evaluated at the bin centred redshifts. While ignoring cross-correlations in the bins we include $\sigma_{i}$ which is the error in $n_{i}$. This error includes Poisson noise, sample variance and the associated redshift uncertainty. In addition to the redshift distribution we also include an HST prior (\citealt{Freedman01}) on $h$ as given by, 

\begin{equation} \label{eq:HSTprior}
\chi^{2}_{\mathrm{HST}} = \frac{1}{2} \frac{(h - 0.72 )^{2}}{0.08^{2}}.
\end{equation}

\begin{figure}
      \includegraphics[width=3.5in,height=3.5in]{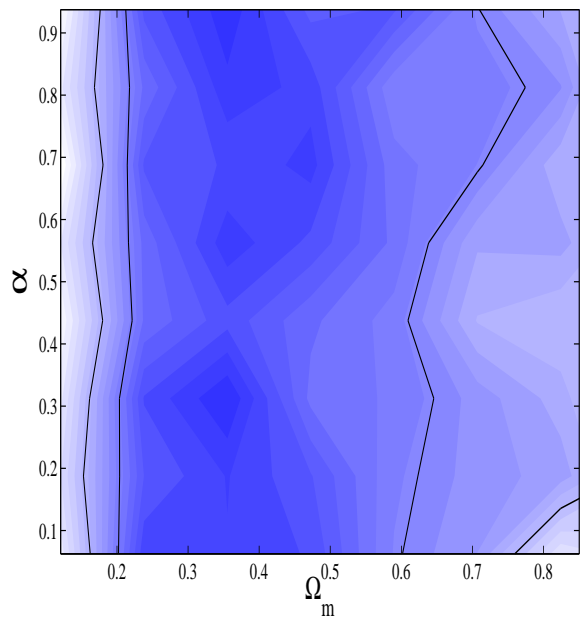}
    \caption{\small{The diagram above demonstrates the attempted constraint on a parameterised gravitational model that is motivated by the concept of a large extra dimension (mDGP). The contours are for $\Omega_{m}$, the matter density, and $\alpha$ the modified gravity parameter where 5 other cosmological parameters ($h$, $\sigma_{8}$, $a$, $b$ and $c$) have been marginalised. Here only angular scales greater than 30 arcminutes have been used in order to avoid the non-linear regime. The selection of data is from the CFHTLS-wide (F08) survey with analysis using the E correlation two point statistic $\xi_{E} + c'$. Note that for this mDGP model $\alpha=0$ corresponds to LCDM whereas $\alpha=1$ is equivalent to the DGP braneworld model.}}
    \label{fig:lensingconstraint}
\end{figure}

For the lensing only analysis we cut the data such that we use angular scales greater than 30 arcminutes, which we reiterate, is to avoid the unknown non-linear contribution to the lensing constraint. The analysis for the mDGP model is shown in Figure \ref{fig:lensingconstraint}. It shows the marginalised $\Omega_{m}$ and $\alpha$ contours where, as shown in Section~\ref{sec:modifiedgravity}, $\alpha$ parameterises a gravitational model with a large extra dimension. The mDGP model interpolates between LCDM ($\alpha = 0$) and the DGP braneworld model ($\alpha=1$). It is clear therefore that a lensing only analysis is presently not capable of constraining mDGP--at least in the context of physically more viable models ($\alpha \lesssim 1$). Indeed we find a similar difficulty in $\gamma$ with no constraint on any reasonable physical values in a lensing only analysis. Firstly, we should perhaps not be too surprised as we have used a relatively new cosmological probe and have, in neglecting the non-linear scales, used only a third of the data points. Tomographic information is not yet currently available for this data either which will vastly improve information on the expansion history and hence $\alpha$. We have also allowed significant cosmological freedom with the variation of 7 parameters. On the other hand, however, this does not mean that lensing, even with a current analysis, is not useful with regards to our late-time acceleration models. In order to see the current effectiveness we now look at Baryon Acoustic Oscillations, Supernovae and lensing in combination in the next section. Then in Section~\ref{sec:Euclid} we see how the future weak lensing survey Euclid will improve upon today's lensing only constraining power. 

\subsection{Supernovae and Baryon Acoustic Oscillations} \label{sec:BAO}

In the previous section we performed a preliminary analysis based on linear to quasi-linear weak lensing data alone. This probe, while having characteristics significant for the discrimation of late time acceleration, was unable to constrain either the mDGP model or the growth parameterisation. As such it is desirable for us to combine it with other cosmological probes in order to improve potential constraints. Moreover, it is most beneficial to combine weak lensing, an indicator of growth and expansion, with distance indicators. This is because of the particular degeneracies that exist between the parameters following an isolated study of lensing. For example, there exists a degeneracy between $w$ and $\gamma$ and so tighter constraints on the expansion history will act to aid any constraint on $\gamma$. Furthermore, inclusion of additional expansion data will aid the constraints of the mDGP model given that different $\alpha$ corresponds to different late time accelerations. We therefore choose to include both Supernovae and Baryon Acoustic Oscillations (BAOs) which, due to their vastly different $\Omega_{m}-w$ degeneracies, are also extremely complementary to one another. 

BAOs, used as standard rulers and observed in the galaxy distribution, aim to test cosmology through the distance-redshift relation. Using the data and notation of \citet{Percival07} we look to utilise the distance measure given by

\begin{equation} \label{eq:BAOdistancemeasure}
D_{V}(z) = [(1+z)^{2} D^{2}_{A} c z / H(z)]^{\frac{1}{3}},
\end{equation}
\noindent
where $D_{A}$ is the angular diameter distance and $H(z)$ is the Hubble parameter. Specifically, it is the ratio $r_{s}/D_{V}(z)$ that we examine where $r_{s}$ is the comoving sound horizon at recombination. \citet{Percival07} detects the BAO in the clustering of 2dFGRS and SDSS galaxy samples and the clustering of SDSS LRGs to quantify this measure at $z=0.2$ and $z=0.35$, respectively. For each likelihood evaluation we compare this data to $r_{s}/D_{V}(z)$ calculated with $D_{V}(z)$ from Equation \eqref{eq:BAOdistancemeasure} and the comoving sound horizon $r_{s}$ evaluated using the formulae in \citet{Eisenstein98}. 

For the inclusion of Supernovae we use the data provided from the first year Supernova Legacy Survey (SNLS) (\citealt{Astier06}). This data set includes 71 type 1a Supernovae also detected at the Canda-France-Hawaii Telescope. Here we use the distance modulus $\mu_{0}$, a measure of the luminosity distance, as the observable.

\begin{equation} \label{eq:supernovaemodulus}
\mu_{0} = 5\mathrm{log}10(d_{L}(z)) + 25.
\end{equation}
\noindent
This is given in the log-likelihood (Equation \eqref{eq:supernovaechisquared}) also as $\mu_{0}$, where $\mu_{B}$ is the observed value. $\sigma_{\mathrm{int}}$ is given by the intrinsic dispersion of the absolute magnitudes and $\sigma(\mu_{B})$ by peculiar velocity and light curve parameter information.

\begin{equation} \label{eq:supernovaechisquared}
\chi^{2} = \sum_{\mathrm{objects}} \frac{( \mu_{B} - 5\mathrm{log}_{10}(d_{L}(\theta,z)) - 25 )^{2}}{\sigma^{2}(\mu_{B}) + \sigma^{2}_{\mathrm{int}}}
\end{equation}

With this machinery in place we are now able to perform additional tests on the mDGP model. We do not constrain the growth parameterisation with these probes in isolation as in this format they have no growth information. We do, however, attempt to constrain the growth with a combined analysis at the end of this section. 

\begin{figure*}
  \begin{flushleft}
    \centering
    \begin{minipage}[c]{1.00\textwidth}
      \centering
      \includegraphics[width=2.75in,height=2.75in]{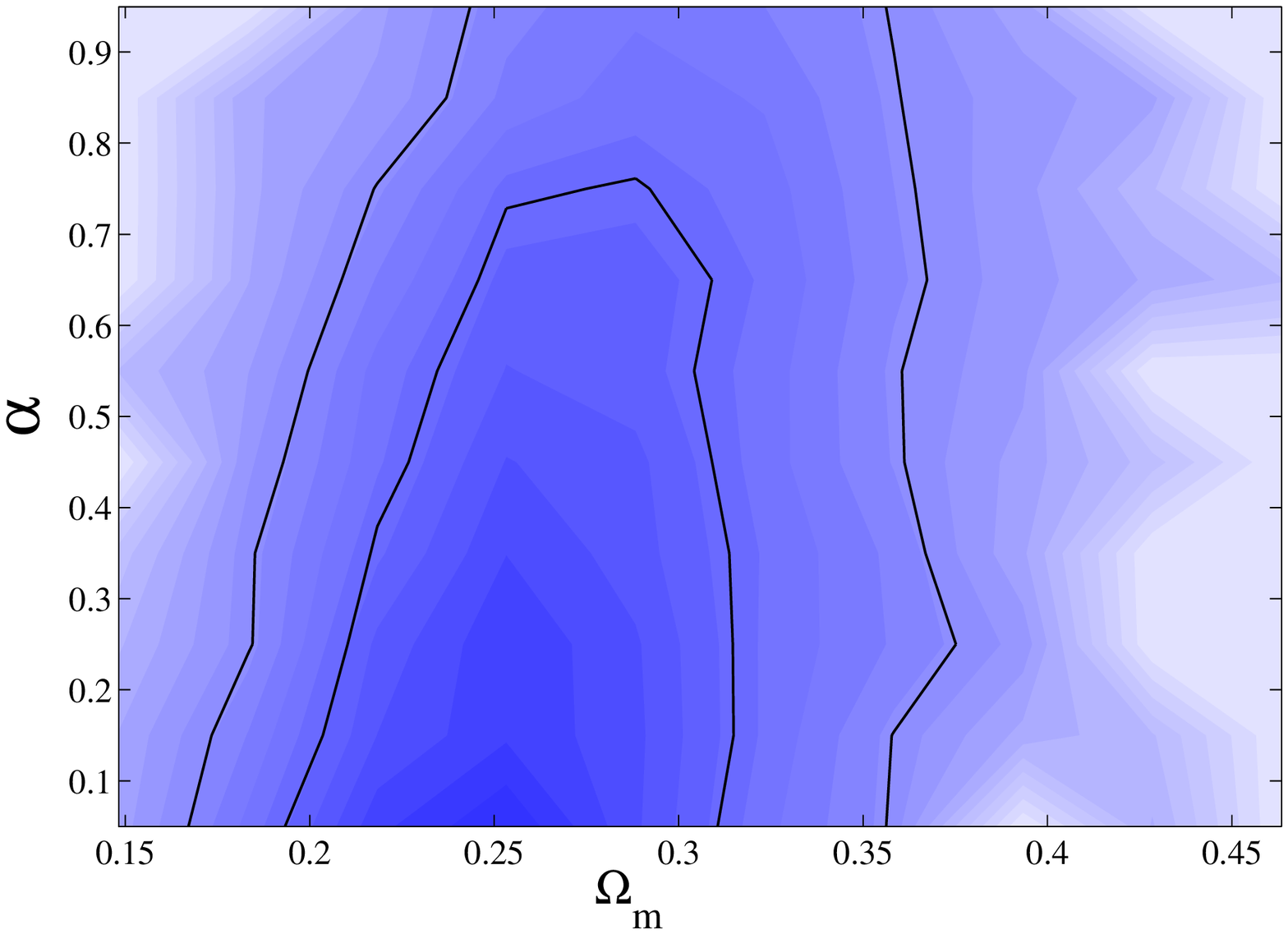} 
      \includegraphics[width=2.75in,height=2.75in]{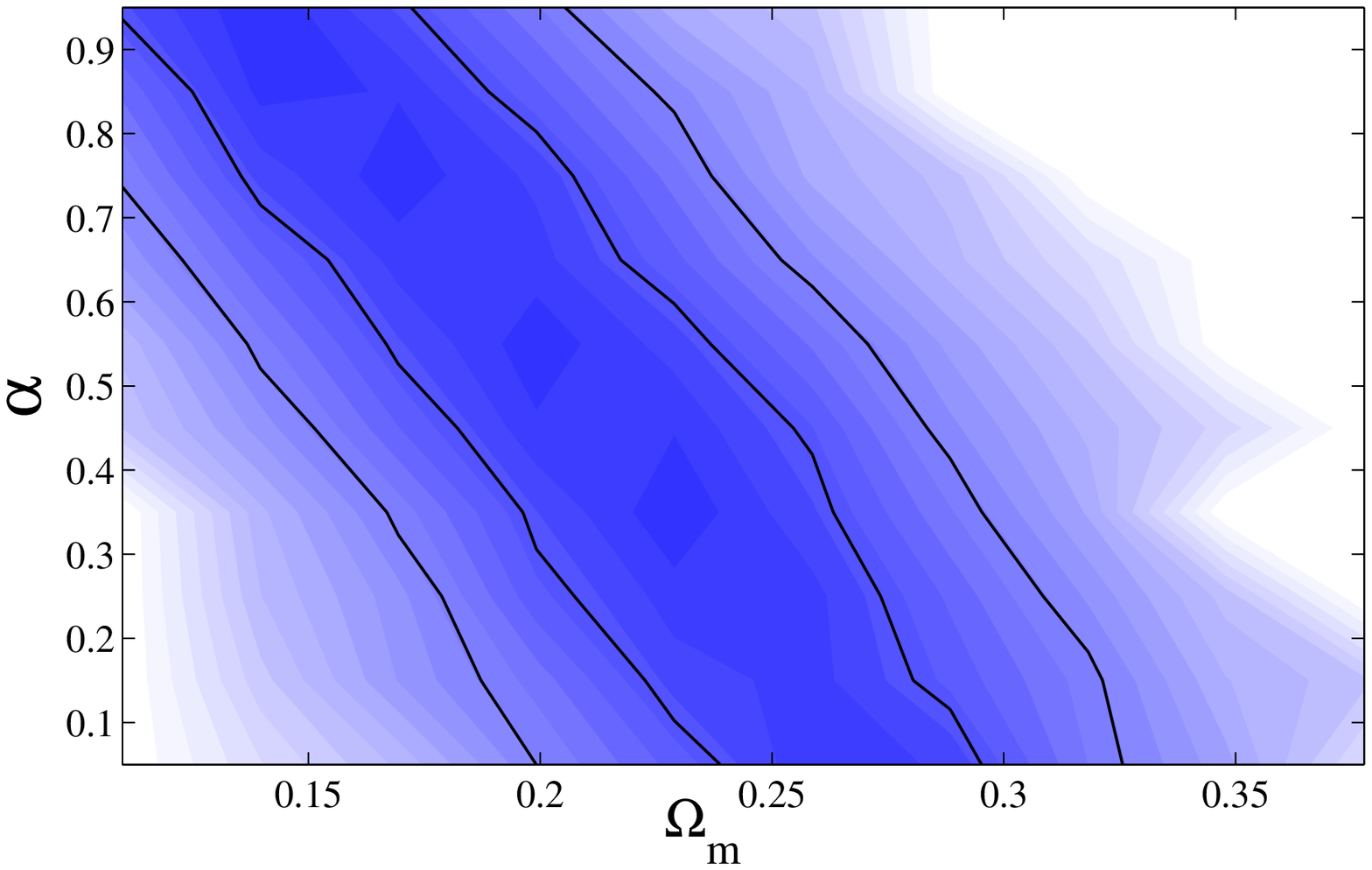} 
    \end{minipage}
    \begin{minipage}[c]{1.00\textwidth}
      \centering
      \includegraphics[width=2.75in,height=2.75in]{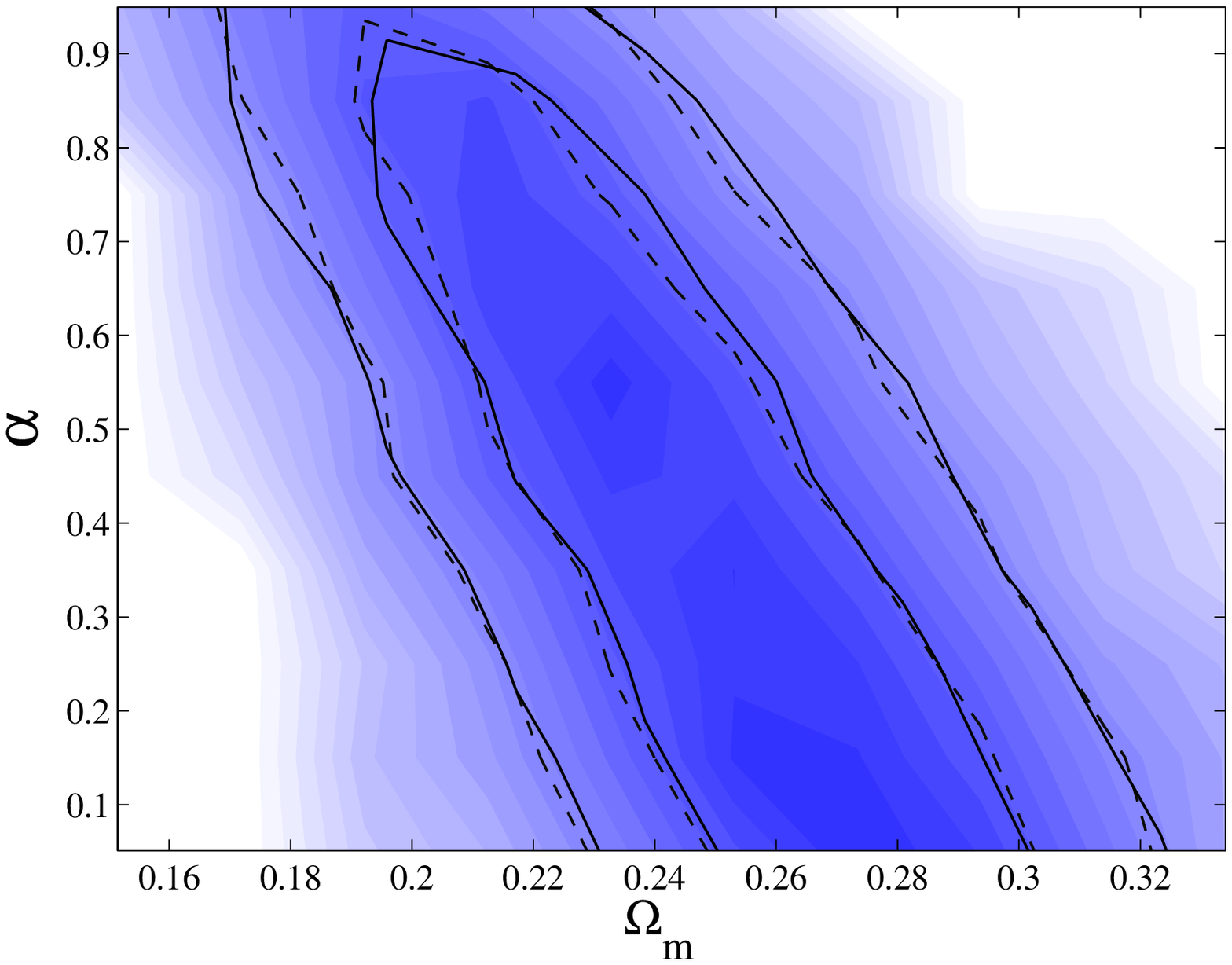} 
      \includegraphics[width=2.75in,height=2.75in]{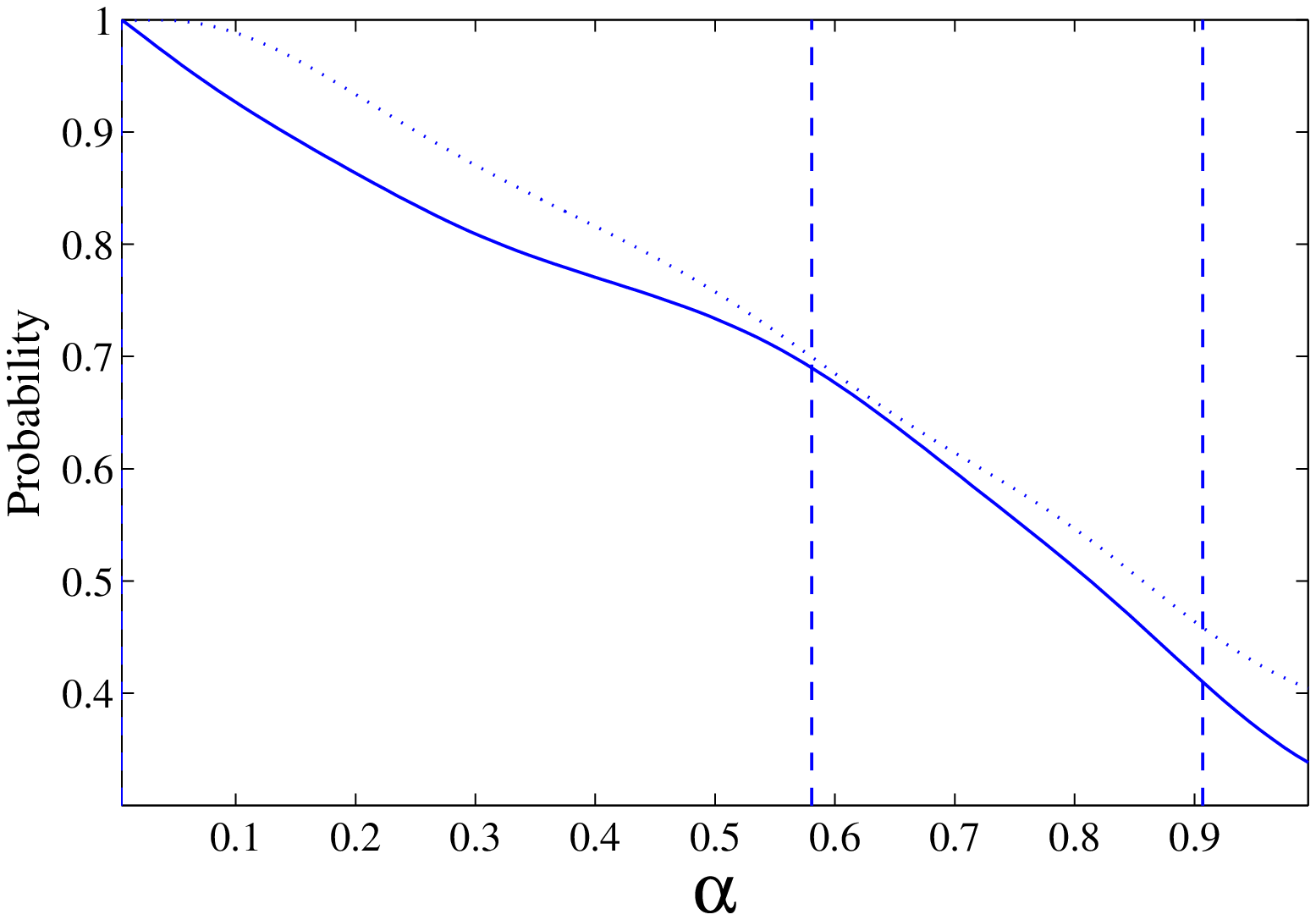} 
    \end{minipage}
    \caption{\small{The plot in the top left panel shows the constraint on $\Omega_{m}$ and $\alpha$. Although appearing to disfavour DGP ($\alpha=1$) as in the analysis by \citet{YamamotoBassettNichol06} the remaining parameters $\Omega_{b}$ and $h_{0}$ have been fixed at 0.044 and 0.66 respectively. We go beyond this in the top right panel which contains constraints given on the same parameters but when Supernovae data is added and $\Omega_{b}$ and $h_0$ are allowed to vary. We see now that the $1\sigma$ contour is beyond the bounds of the plot and so no constraint can be inferred. The benefit of the weak lensing data is seen in the bottom left panel where once again we use angular scales greater than 30 arcminutes from the CFHTLS-wide (F08) lensing survey. We also vary $\Omega_{m}$, $h$, $\sigma_{8}$, $\Omega_{b}$, $a$, $b$, $c$ and $\alpha$ whilst keeping $n_{s} = 0.963$. With this addition it is evident that there is a visible improvement in constraint and that DGP is marginally disfavoured. This is exemplified in the bottom right panel where we include the 1D marginalised probability distribution (solid line). We find that the joint analysis gives constraints on mDGP of $\alpha < 0.58$ and $\alpha < 0.91$ at the $68\%$ and $95\%$ confidence levels, respectively. The dotted line represents the mean likelihood of the samples. Finally, the dashed contours in the bottom left hand panel show that our constraints are roughly insensitive to any systematics in the data such as an over or underestimation of the shear at high redshift (Section~\ref{sec:systematics}).}}
    \label{fig:BAOandSupernovae}
  \end{flushleft}
\end{figure*}

\begin{figure*}
  \begin{flushleft}
    \centering
    \begin{tabular}{ll}
      \includegraphics[width=3.25in,height=3.25in]{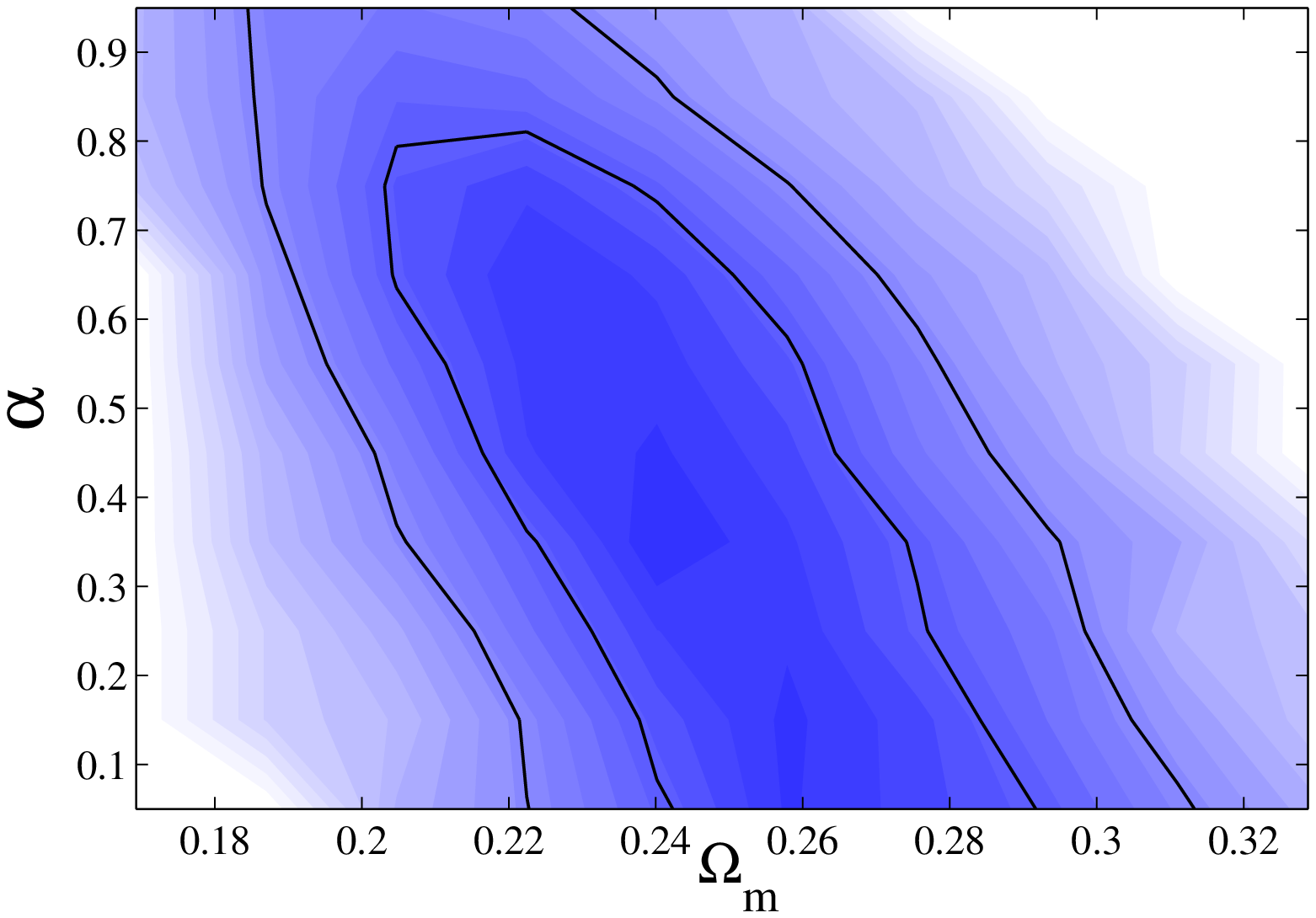} &
      \includegraphics[width=3.25in,height=3.25in]{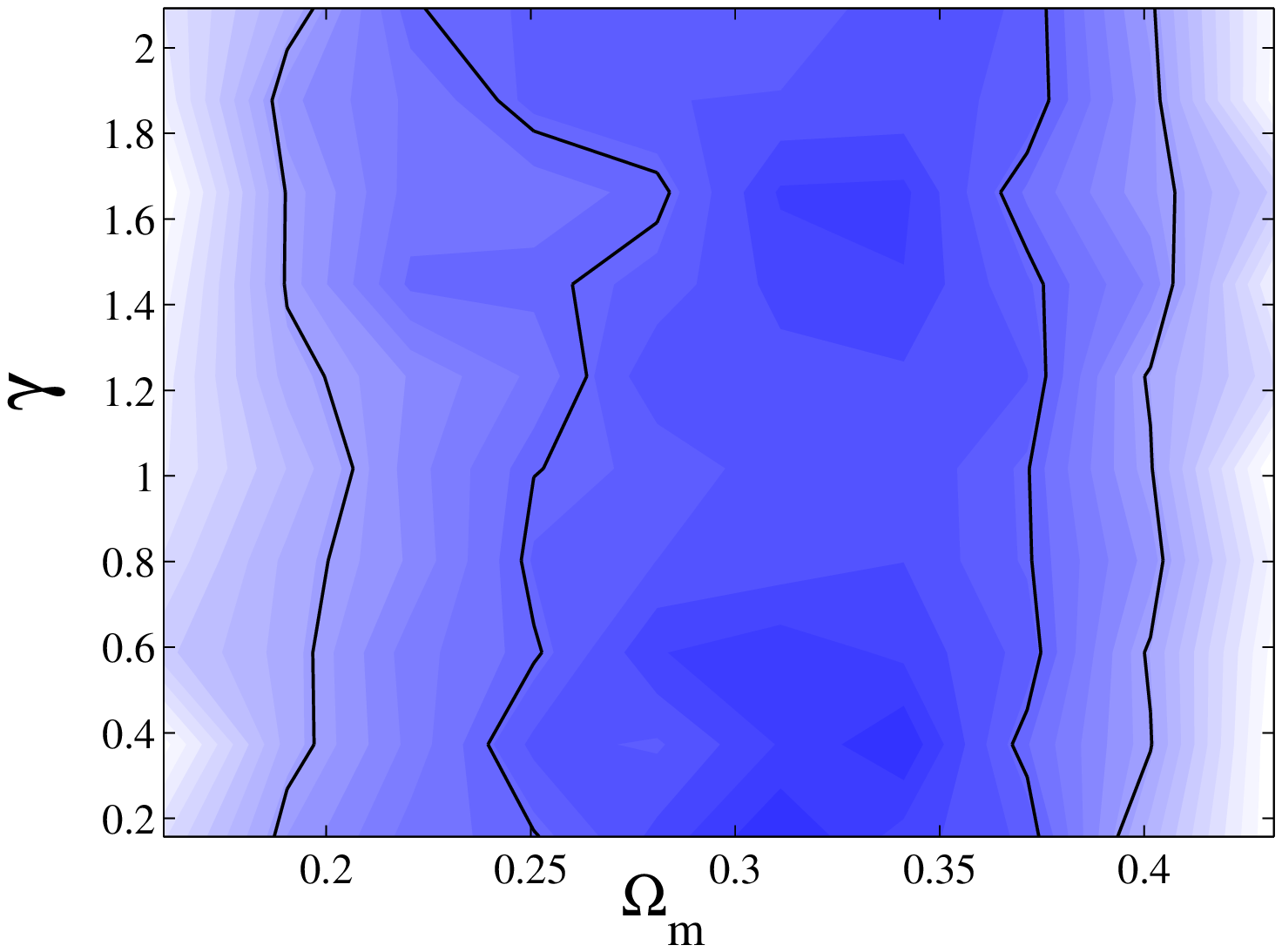}
    \end{tabular}
    \caption{\small{The left panel is an analysis of the mDGP model with weak lensing, BAO and Supernovae, as before, but in this case with the full range of angular scales (1-230 arcminutes). There is a slight, but not significant, improvement than in the more linear analysis. Here we find $\alpha < 0.56$ and $\alpha < 0.86$ at the $68\%$ and $95\%$ confidence levels, respectively. It should be noted that this analysis includes data from the unknown non-linear regime however. The right panel demonstrates the current challenge in constraining the gravitational--as opposed to the expansion--contribution to the growth of structure. We find that with current data we are unable to put any bounds on reasonable values of the $\gamma$ parameter. This plot contains an analysis with weak lensing, Supernovae and BAOs. Implicit in this plot is the variation of also $h$, $\sigma_{8}$, $\Omega_{b}$, $a$, $b$, $c$ and $w_{0}$. Only angular scales greater than 30 arcminutes have been used.}}
    \label{fig:combinedprobes}
  \end{flushleft}
\end{figure*}

By looking at the top left hand panel of Figure \ref{fig:BAOandSupernovae} we can see that it is possible to place a constraint on $\alpha$ with a BAO only analysis. It should be noted that the apparent disfavouring of DGP ($\alpha=1$) is not as promising as it first appears for we have, replicating the analysis of \citet{YamamotoBassettNichol06} with BAO only, varied just $\Omega_{m}$ and $\alpha$ with $\Omega_{b}$ and $h$ held fixed at 0.044 and 0.66, respectively. The top right hand panel of the same Figure demonstrates the need for caution as we go beyond the \citet{YamamotoBassettNichol06} analysis. By allowing more cosmological freedom (i.e. varying $\Omega_{b}$ and $h$) and including the Supernovae data the two probes are in fact incapable of disfavouring DGP. We can now see how, even at present, the weak lensing data is useful for the study of this modified gravity. The $\Omega_{m}-\alpha$ contours for a joint analysis with all three combined probes is displayed in the bottom left hand panel. Once again we use only angular scales greater than 30 arcminutes to avoid the unknown non-linear regime. For the combined analysis it is worth noting that we vary a significant total of 8 cosmological parameters: $\Omega_{m}$, $h$, $\sigma_{8}$, $\Omega_{b}$, $a$, $b$, $c$ and $\alpha$. It is now evident that the addition of the weak lensing analysis is beneficial with a mild disfavouring of the DGP end of the $\alpha$ spectrum. Indeed, we include in the bottom right hand panel the 1D probability distribution for $\alpha$, in the process demonstrating that $\alpha < 0.58$ at the $68\%$ confidence level and $\alpha < 0.91$ at the $95\%$ confidence level. This corresponds to a disfavouring of DGP at over $2\sigma$. Furthermore, we include in the left panel of Figure \ref{fig:combinedprobes}, for interest, the same analysis but with all angular scales (1-230 arcminutes). There is a noticeable but not significant improvement in the constraint on mDGP. The slight increase in disfavouring leads to $\alpha < 0.56$ and $\alpha < 0.86$ at the $68\%$ and $95\%$ confidence levels, respectively. Again to reiterate it should be noted that this analysis, in using all angular scales, ventures into the non-linear regime.

Having had some success with a combination of the three cosmological probes it is worth investigating whether they can aid the determination of the far more subtle growth parameterisation $\gamma$. While the BAO and Supernova will not add growth information explicitely they may help reduce the parameter degeneracies and cut the parameter space favourably. We find however, by looking at the right panel of Figure \ref{fig:combinedprobes}, that at present and for meaningful values of $\gamma$ we still have insufficient constraining power. This is understandable given that for the mDGP model $\alpha$ contained growth and expansion information. In this way the Supernovae and the BAO actively contributed to the constraint while the lensing constrained it through the comoving diameter distances $D(\chi)$ in the convergence power spectrum, through expansion terms within the growth via the Hubble drag and finally through the pure growth contribution as seen in the $\beta$ contribution. Constraining $\gamma$ on the other hand is equivalent to just changes in $\beta$ and is therefore far subtler. However, just because we do not have the current data to pick out this effect it should not deter us from pursuing this potential signature of modified gravity. In fact, future cosmological probes, such as weak lensing, may well be able to extract this contribution and tighten constraints on beyond-Einstein cosmology. We look to the future in Section~\ref{sec:Euclid}.

\subsection{Accounting for systematics} \label{sec:systematics}

Despite the absence of any significant B-modes in the CFHTLS data there exists a potential underestimation of the shear at high redshift (Kilbinger et al. in prep.). In order to account for this potential effect we use the model introduced in Kilbinger et al. and multiply the redshift distribution at high redshift ($z>1$) by some constant $c_{0}$--where $c_{0} < 1$ constitutes underestimation, $c_{0} > 1$ overestimation and $c_{0} = 1$ no alteration in the shear. Following the aforementioned paper we vary $c_{0}$ and marginalise over the parameter in an additional cosmological run, thus accounting for any such systematic. We subsequently find little change in the constraint on mDGP as shown by the dashed contour in the bottom left hand panel of Figure \ref{fig:BAOandSupernovae}. Indeed the $68\%$ confidence interval is only shifted from $\alpha < 0.58$ to $\alpha < 0.61$ with systematic treatment. Moreover, with the cut in angular scales at $\theta = 30$ arcminutes it has not been possible to meaningfully constrain $c_{0}$ itself. Further treatment and causes of potential systematics in the data are detailed in van Waerbeke et al. in prep. and Kilbinger et al. in prep.

\section{Future Probes - Euclid} \label{sec:Euclid}

Within the foreseeable future the age of precision cosmology looks likely to get ever more precise. The field of weak lensing is most definitely no exception and it is perhaps set to be one of the most promising areas of development. One striking reason, among others, is the planned Euclid mission (\citet{Refregier08} and \citet{Cimatti08}). This space based project aims to have an effective sky coverage of $20000$ square degrees, probing the galaxy distribution with a median redshift $z_{m} = 0.9$ and giving an effective galaxy density of $40$ $\mathrm{gal/arcmin^{2}}$. Hence it is of great interest to see how this future survey will probe the nature of gravity, whatever that may be. We therefore undertake a Fisher matrix analysis with the intention of forecasting, with lensing only, how Euclid will be able to constrain the mDGP model ($\alpha$) as well as the general parameterisation for gravity ($\gamma$ and $\Sigma$). The Fisher matrix formalism for weak lensing is given by

\begin{equation} \label{eq:weaklensingfisher}
F_{ij} = \sum_{l} \frac{\partial C}{\partial p_{i}} \mathrm{Cov}^{-1} \frac{\partial C}{\partial p_{j}},
\end{equation}
\noindent
where $C$ is the weak lensing observable shown by
\begin{equation} \label{eq:observedlensingpowerspectrum}
C_{ij}(l) = P^{k}_{ij} + <\! \gamma_{\mathrm{int}}^{2} \!> \delta_{ij}/\bar{n_{i}},
\end{equation}
\noindent
and $P^{k}_{ij}$ is the power spectrum similar to Equation \eqref{eq:convergencepowerspectrum} but with indices i and j denoting tomographic bins. Further still, $\bar{n_{i}}$ is the average galaxy number per steradian in bin $i$ and $< \! \gamma_{\mathrm{int}}^{2} \! >^{\frac{1}{2}}$ is the rms intrinsic shear in each component, which is equal here to $0.22$. $\partial p_{i}$ denotes the derivative with respect to a parameter $p$ that is varied and Cov is the covariance matrix given by,

\begin{equation} \label{eq:covariance}
\mathrm{Cov}[C^{k}_{ij}(l),C^{k}_{kl}(l')] = \frac{\delta_{ll'}}{(2l+1)f_{\mathrm{sky}} \Delta l} [C^{k}_{ik}(l) C^{k}_{jl}(l) + C^{k}_{il}(l)C^{k}_{jk}(l)].
\end{equation}

For our analysis we assume the redshift distribution given in Equation \eqref{eq:n_z} where $z_{0} = z_{m}/1.412$, $\alpha = 2$ and $\beta=1.5$. We allow for five redshift bins with divisions such as to give approximately equal galaxy number in each bin. The photometric error is also accounted for by using the parameterisation $\sigma_{z}= \sigma_{p}(1+z)$, with $\sigma_{p}$ taken to be $0.03$.

\begin{equation} \label{eq:n_z}
n(z) \propto z^{\alpha} \mathrm{exp}(-(z/z_{0})^{\beta})
\end{equation}

For the base cosmology we vary, about their fiducial values, the parameters $\Omega_{\mathrm{m}}=0.3$, $h=0.7$, $\sigma_{8}=0.8$, $\Omega_{\mathrm{b}}=0.05$ and $n_{s}=1.0$, which are common to both analyses, with also $w_{0}=-1.0$, $w_{a}=0.0$, $\gamma = 0.55$ and $\Sigma_{0} = 0$ (see Section~\ref{sec:growth}) for the general parameterisation, and $\alpha = 0$ for the mDGP model. We therefore vary a full 9 parameters for the general modifed gravity parameterisation and 6 parameters for the specific braneworld model--remembering that $\alpha$ specifies its own $w_{0}$, $w_{a}$, $\gamma$ and $\Sigma_{0} = 0$.

\begin{figure*}
  \begin{flushleft}
    \centering
    \begin{tabular}{ll}
      \includegraphics[width=3.25in,height=3.25in]{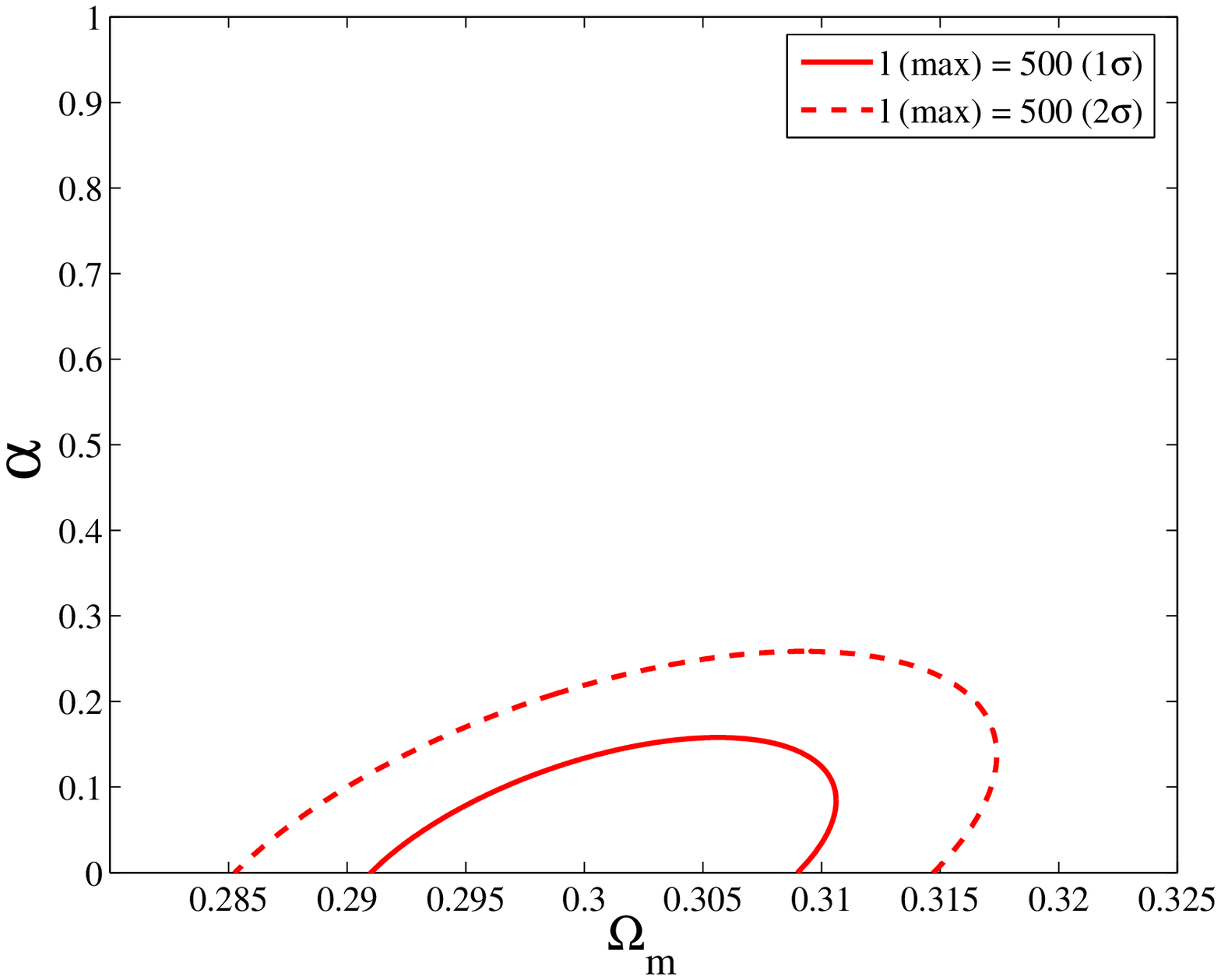} &
      \includegraphics[width=3.25in,height=3.25in]{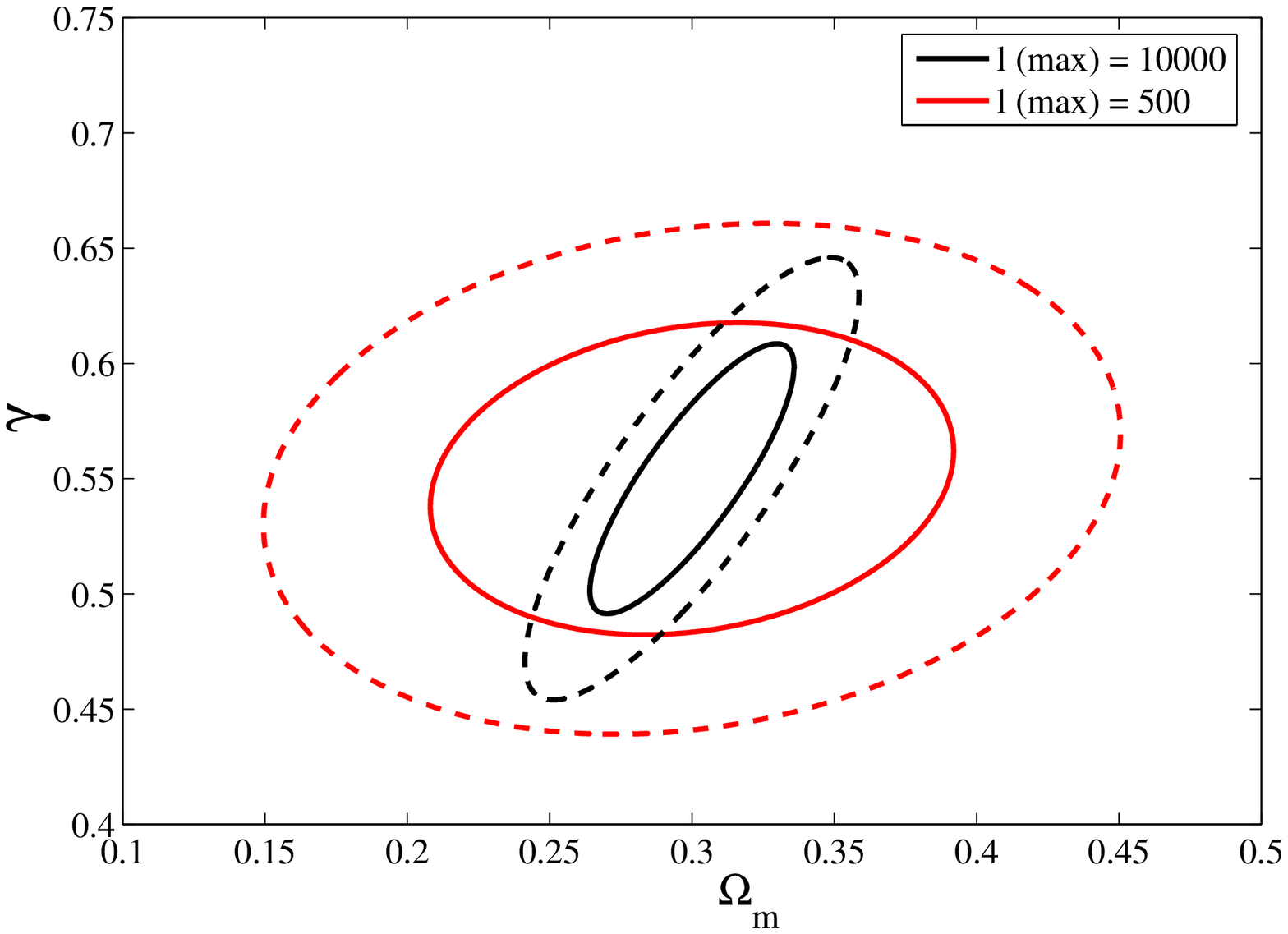}
    \end{tabular}
    \caption{\small{The left panel displays the ability of Euclid's potential constraining power with regards to the mDGP model in a lensing only analysis. Here the $1\sigma$ contour (all solid lines) is well within the $\alpha=1$, or DGP, line and so it will be easily distinguishable from LCDM ($\alpha=0$). Indeed this corresponds to an error of 0.104 on $\alpha$ with $l_{\mathrm{max}}=500$ (all red contours) in stark contrast to today's constraint. The right panel shows the marginalised contours for the general growth parameterisation. Again, it seems that Euclid will provide excellent insight into any potential modified gravity signatures. Specifically it is found that it will be possible to constrain $\gamma$ with an error of 0.045 ($1\sigma$). This is tightened further to an error of 0.038 when $l_{\mathrm{max}}=10000$ (black contours). The parameters $h$, $\sigma_{8}$, $\Omega_{\mathrm{b}}$ and $n_{s}$ have been varied and marginalised over for both models considered here while in addition $w_{0}$, $w_{a}$ and $\Sigma_{0}$ have been marginalised for the growth model.}}
    \label{fig:Euclidforecast}
  \end{flushleft}
\end{figure*}

The resulting forecasts for this future project can be seen clearly in Figure \ref{fig:Euclidforecast}. The left panel shows that Euclid will be able to put considerable strain on any braneworld-like gravity scenario that resembles the mDGP model. The solid ($1\sigma$) and dashed ($2\sigma$) contours are significantly within the $\alpha=1$, or DGP, bound. In fact, this demonstrates that Euclid will potentially constrain $\alpha$ to within an error of 0.104 at the $68\%$ confidence level. This is in constrast to Figure \ref{fig:lensingconstraint} where no constraint was possible with a lensing-only study. Indeed, note that for this analysis only contributions from $l=10$ to $l_{\mathrm{max}} = 500$ were considered such that the deeply non-linear regime could be neglected.

The right panel in Figure \ref{fig:Euclidforecast} again illustrates the expected constraining power of Euclid but now with regards to general modified gravity. For this general parameterisation we performed two runs with one corresponding to l contributions from $l=10$ to $l_{\mathrm{max}} = 500$ (red contours) and the other with contributions from $l=10$ to $l_{\mathrm{max}} = 10000$ (black contours). Here we see that Euclid will be able to extract the growth characteristic thus allowing, potentially, a strong cosmological test of our gravitational framework. Indeed, it will be able to constrain the fiducial $\gamma =0.55$ (LCDM) to within an error of 0.0446 at $1\sigma$ with $l_{\mathrm{max}} = 500$ which is further tightened to 0.038 with $l_{\mathrm{max}} = 10000$. The other 7 cosmological parameters ($h$, $\sigma_{8}$, $\Omega_{\mathrm{b}}$, $w_{0}$, $w_{a}$, $n_{s}$ and $\Sigma_{0}$) have been varied and marginalised over. Again this forecast is in contrast to Figure \ref{fig:combinedprobes} where even the combined probes of lensing, BAOs and Supernovae were incapable of constraining the growth.

\begin{figure}
      \includegraphics[width=3.5in,height=3.5in]{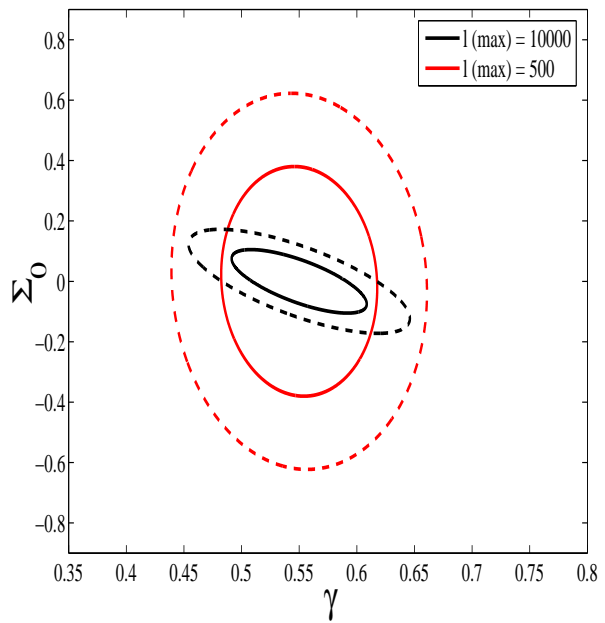}
    \caption{\small{The above plot shows the marginalised $\gamma-\Sigma_{0}$ forecast for a weak lensing only analysis with Euclid. These two parameters, which could represent modified gravity or generic dark energy signatures, demonstrate how this future weak lensing probe will potentially place firm constrains on any model of late-time acceleration. The black contours correspond to $l_{\mathrm{max}} = 10000$, demonstrating an error of 0.069($1\sigma$) on $\Sigma_{0}$, whereas the red contours correspond to $l_{\mathrm{max}} = 500$ giving instead an error of 0.25. In both cases the inner and outer contours are $1\sigma$ and $2\sigma$, respectively.}}
    \label{fig:modified}
\end{figure}

Finally, we include in Figure \ref{fig:modified} a marginalised contour for $\gamma$ against $\Sigma_{0}$ which further highlights how modified gravity or very generic dark energy signatures can be constrained, consistently, with weak lensing. We find that with $l_{\mathrm{max}} = 500$ (red contours) this survey could constrain alterations in the powerspectrum with an error in $\Sigma_{0}$ of 0.25 at $1\sigma$. This parameter is more sensitive to the range of scales used than $\gamma$, however, with $l_{\mathrm{max}} = 10000$ (black contours) confining $\Sigma_{0}$ to within 0.069 of the fiducial value $\Sigma_{0} = 0$. 

We should emphasise that the spectroscopic element of Euclid \citep{Cimatti08} will also be able to constrain growth via redshift space distortions (\citet{Peacock02} and \citet{Guzzo08}).

\section{Discussion and Conclusion} \label{sec:conclusion}

To summarise, we have noted that the surprising, but well confirmed, late-time acceleration of the universe \emph{could} be the result of a modification to gravity. We then, in Section~\ref{sec:modifiedgravity}, reviewed the concept of modified gravity detailing in the process a model that is motivated and acts to parameterise a large extra dimension (mDGP). Interpolating between LCDM ($\alpha=0$) and DGP ($\alpha=1$) it can be tested as a model in its own right and/or used as an example to demonstrate the rich set of observational signatures that are likely to arise for a modified gravity model in cosmology. These signatures include the expansion history, and rather interestingly, the growth of structure and a modification to the relationship between the potential power spectrum and the matter power spectrum. With these characteristics in mind we then discussed attempts in the literature to parameterise modified gravity in this way. This included a growth parameter $\gamma$ and a power spectrum parameter $\Sigma$. 

In Section~\ref{sec:weaklensingasacosmologicalprobe} we introduced weak lensing and related its attributes to modified gravity given that it is sensitive to the expansion history, the growth of structure and the power spectrum. We noted a severe caveat in the use of non-linear scales and therefore, in Section~\ref{sec:CFHTLS}, described the approapriate choice of survey (CFHTLS-wide) and data ($\theta > 30$ arcminutes) used in the cosmological analyses. The subsequent lensing only constraints were given in Section~\ref{sec:lensing} where we showed that one could not yet constrain meaningful values of $\alpha$ or $\gamma$ with the current data. We then, by adding it to Baryon Acoustic Oscillation and Supernovae data, demonstrated that weak lensing was highly beneficial in aiding the constraint of mDGP in combination. For without the inclusion of the lensing data the expansion-only probes were incapable of constraining $\alpha=1$. We found however that the combined probes disfavoured the DGP model with over a $95\%$ confidence level where specifically $\alpha < 0.58$ at $1\sigma$ and $\alpha < 0.91$ at $2\sigma$. We then showed that a constraint on the subtle, yet potentially important, growth signature is beyond the current weak lensing, BAO and Supernovae data. We allowed almost total cosmological freedom in all these analyses, varying parameters: $\Omega_{m}$, $h$, $\sigma_{8}$, $\Omega_{\mathrm{b}}$, $a$, $b$ and $c$ for both models, in addition to $\alpha$ for mDGP and $w_{0}$ and $\gamma$ for the growth model. Furthermore, we used the $\xi_{E}$ two point statistic while analytically marginalising over the residual offset $c'$. We also demonstrated in Section~\ref{sec:systematics} that our results are insensitive to a potential over or underestimation of the shear at high redshift.

Finally in Section~\ref{sec:Euclid} we looked towards the future space based weak lensing survey, Euclid, and discovered that it will have significant ability to distinguish between modified gravity and LCDM. We included a forecast for the mDGP model finding that even for a lensing only analysis Euclid could restrict $\alpha$ to within $0.104$ of the fiducial $\alpha = 0$ at $1\sigma$, even when the deeply non-linear regime has been removed ($l_{\mathrm{max}} = 500$). In addition, we included a complete and consistent forecast for generalised modified gravity demonstrating that deviations from a fiducial $\Sigma_{0} = 0$ of $0.25$ at the $68\%$ confidence level will be possible with $l_{\mathrm{max}} = 500$. With $l_{\mathrm{max}} = 10000$ this gets restricted to  $\Delta \Sigma_{0} = 0.069$. It will also confine $\gamma$ to within $0.045$ ($l_{\mathrm{max}} = 500$) and $0.038$ ($l_{\mathrm{max}} = 10000$) of the fiducial $\gamma = 0.55$ at $1\sigma$, where a full 9 cosmological parameters were varied.

In our analyses with data we have, except as an example case, actively removed angular scales less than 30 arcminutes. This was to remove the contribution from the unknown non-linear regime from the constraints. This constitutes removing vast quantities of attainable data. Not only is this clearly not maximising the available information but it could be that non-linear physics acts to emphasise any difference in gravitational theory. Therefore it would naturally be highly beneficial to understand the evolution of these density perturbations within this regime. In order to do this there needs to be continuing attempts in the community to study N-body simulations with varying gravitational frameworks.

Even though we have detailed the advantages of weak lensing in a modified gravity study, and even though Euclid in particular will be deeply insightful it seems, however, a collected and coordinated assault on our gravitational framework would prove even more advantageous. This might exist in the form of a combination of probes as, for example, discussed in \citet{Jain07}, where ideally the four perturbation variables $\phi$, $\psi$, $\delta$ and $\theta$ are independently targeted. This in principle would allow us unprecidented experimental scrutiny on the structure of our gravitational theory over large scales. However, to conclude, it is clear that to actually solve the tantalising problem that grips cosmology there needs to be a paradigm shift analogous to those that have revealed the great problems of the past.
\\
\\

Acknowledgements: It is a pleasure to thank the CFHTLS team for their kind distribution of the lensing data. Particular thanks go to Martin Kilbinger and also Liping Fu for answering and discussing some of the more technical aspects of the data and potential systematics. We would also like to thank the Euclid internal referee for careful reading of this script. ST would like to thank Adam Amara for thorough and detailed weak lensing and fisher matrix code comparisons. In addition, ST would like to thank Sarah Bridle for producing and distributing CosmoloGUI which was used to produce the contours in this paper. Finally, FBA acknowledges the Leverhulme Foundation for support through an Early Careers Fellowship.

\appendix

\bibliographystyle{./mn2e.bst}

\bibliography{./aamnem99,./references}

\end{document}